%% file: agpaz.tex
\documentclass[12pt,preprint]{aastex} 

\newcommand{\noprint}[1]{} 
\newcommand{\figsetstart}{{\bf Fig. Set} } 
\newcommand{\figsetend}{} 
\newcommand{\figsetgrpstart}{}
\newcommand{\figsetgrpend}{}
\newcommand{\figsetnum}[1]{{\bf #1.}}
\newcommand{\figsettitle}[1]{ {\bf #1} }
\newcommand{\figsetgrpnum}[1]{\noprint{#1}}
\newcommand{\figsetgrptitle}[1]{\noprint{#1}}
\newcommand{\figsetplot}[1]{\noprint{#1}}
\newcommand{\figsetgrpnote}[1]{\noprint{#1}}

\shorttitle{GALEX UV Atlas of Nearby Galaxies}
\shortauthors{Gil de Paz et al.}

\begin{document}

\title{The GALEX Ultraviolet Atlas of Nearby Galaxies}

\author{Armando Gil de Paz\altaffilmark{1}, Samuel Boissier\altaffilmark{1}, Barry F. Madore\altaffilmark{1,2}, Mark Seibert\altaffilmark{3}, Young H. Joe\altaffilmark{1,4}, Alessandro Boselli\altaffilmark{5}, Ted K. Wyder\altaffilmark{3}, David Thilker\altaffilmark{6}, Luciana Bianchi\altaffilmark{6}, Soo-Chang Rey\altaffilmark{3}, R. Michael Rich\altaffilmark{7}, Tom A. Barlow\altaffilmark{3}, Tim Conrow\altaffilmark{3}, Karl Forster\altaffilmark{3}, Peter G. Friedman\altaffilmark{3}, D. Christopher Martin\altaffilmark{3}, Patrick Morrissey\altaffilmark{3}, Susan G. Neff\altaffilmark{8}, David Schiminovich\altaffilmark{3}, Todd Small\altaffilmark{3}, Jos\'{e} Donas\altaffilmark{5}, Timothy M. Heckman\altaffilmark{9}, Young-Wook Lee\altaffilmark{4}, Bruno Milliard\altaffilmark{5}, Alex S. Szalay\altaffilmark{9}, Sukyoung Yi\altaffilmark{4}}

\altaffiltext{1} {The Observatories, Carnegie Institution of Washington, 813 Santa Barbara Street, Pasadena, CA 91101; agpaz, boissier, young@ociw.edu; AGdP's current address: Departamento de Astrof\'{\i}sica, Universidad Complutense de Madrid, Madrid 28040, Spain; SB's current address: Laboratoire d'Astrophysique de Marseille, BP 8, Traverse du Siphon, 13376 Marseille Cedex 12, France.}
\altaffiltext{2} {NASA/IPAC Extragalactic Database, California Institute of Technology, MS 100-22, Pasadena, CA 91125; barry@ipac.caltech.edu} 
\altaffiltext{3} {California Institute of Technology, MC 405-47, 1200 East California Boulevard, Pasadena, CA 91125; mseibert, wyder, screy, tab, tim, krl, friedman, cmartin, patrick, ds, tas@srl.caltech.edu; SCR's current address: Department of Astronomy and Space Sciences, Chungnam National University, 220 Gung-dong, Yuseong-gu, Daejeon 305-764, Korea}
\altaffiltext{4}{Center for Space Astrophysics, Yonsei University, Seoul 120-749, Korea;  ywlee@csa.yonsei.ac.kr, yi@astro.ox.ac.uk}
\altaffiltext{5}{Laboratoire d'Astrophysique de Marseille, BP 8, Traverse du Siphon, 13376 Marseille Cedex 12, France; alessandro.boselli, jose.donas, bruno.milliard@oamp.fr}
\altaffiltext{6}{Center for Astrophysical Sciences, The Johns Hopkins University, 3400 N. Charles St., Baltimore, MD 21218; dthilker, bianchi@pha.jhu.edu}
\altaffiltext{7}{Department of Physics and Astronomy, University of California, Los Angeles, CA 90095; rmr@astro.ucla.edu}
\altaffiltext{8}{Laboratory for Astronomy and Solar Physics, NASA Goddard Space Flight Center, Greenbelt, MD 20771; neff@stars.gsfc.nasa.gov}
\altaffiltext{9}{Department of Physics and Astronomy, The Johns Hopkins University, Homewood Campus, Baltimore, MD 21218; heckman, szalay@pha.jhu.edu}

\begin{abstract}
We present images, integrated photometry, surface-brightness and color
profiles for a total of 1034 nearby galaxies recently observed by the
Galaxy Evolution Explorer (GALEX) satellite in its far-ultraviolet
(FUV; $\lambda_{\mathrm{eff}}$=1516\,\AA) and near-ultraviolet (NUV;
$\lambda_{\mathrm{eff}}$=2267\,\AA) bands. Our catalog of objects is
derived primarily from the GALEX Nearby Galaxies Survey (NGS)
supplemented by galaxies larger than 1\,arcmin in diameter
serendipitously found in these fields and in other GALEX exposures of
similar of greater depth. The sample analyzed here adequately
describes the distribution and full range of properties (luminosity,
color, Star Formation Rate; SFR) of galaxies in the Local Universe.

From the surface brightness profiles obtained we have computed
asymptotic magnitudes, colors, and luminosities, along with the
concentration indices C31 and C42. We have also morphologically
classified the UV surface brightness profiles according to their
shape. This data set has been complemented with archival optical,
near-infrared, and far-infrared fluxes and colors. 


We find that the integrated (FUV$-K$) color provides robust
discrimination between elliptical and spiral/irregular galaxies and
also among spiral galaxies of different sub-types. Elliptical galaxies
with brighter $K$-band luminosities (i.e$.$ more massive) are redder
in (NUV$-K$) color but bluer in (FUV$-$NUV) (a color sensitive to the
presence of a strong UV upturn) than less massive ellipticals. In the
case of the spiral/irregular galaxies our analysis shows the presence
of a relatively tight correlation between the (FUV$-$NUV) color (or,
equivalently, the slope of the UV spectrum, $\beta$) and the total
infrared-to-UV ratio. The correlation found between (FUV$-$NUV) color
and $K$-band luminosity (with lower luminosity objects being bluer
than more luminous ones) can be explained as due to an increase in the
dust content with galaxy luminosity.

The images in this Atlas along with the profiles and integrated
properties are publicly available through a dedicated web page at {\tt
http://nedwww.ipac.caltech.edu/level5/GALEX\_Atlas/}
\end{abstract}


\section{Introduction}
\label{introduction}

There are several compelling reasons for observing nearby galaxies in
the ultraviolet (UV). First of all, massive, young stars emit most of
their energy in this part spectrum and at least in star-forming
galaxies they outshine the emission from any other stage of the
evolution of a composite stellar population (e.g$.$ Bruzual \& Charlot
2003). Therefore, the flux emitted in the UV in spiral and irregular
is an excellent measure of the current Star Formation Rate (SFR;
Kennicutt 1998; Donas et al$.$ 1987).  In the case of quiescent
elliptical galaxies the analysis of the UV upturn (the rising part of
the FUV spectrum of these galaxies) promises to provide fundamental
clues in our understanding of the evolution of low-mass stars on the
horizontal branch. Due to its remarkable sensitivity to the physical
properties of these stars, the UV upturn could be used, once fully
understood, as a powerful diagnostic of old stellar populations
(Burstein et al$.$ 1988; O'Connell 1999; Yi et al$.$ 1999; Brown 2004;
Rich et al$.$ 2005, 2006, in prep$.$; Boselli et al$.$ 2005). The UV
has also revealed the presence of residual star formation in a
non-negligible fraction of low-redshift elliptical galaxies (Yi et
al$.$ 2005).

Second, the light emitted in the UV can be very efficiently absorbed
by dust and then re-emitted at far-infrared (FIR)
wavelengths. Therefore an analysis of the energy budget using a
comparison of the infrared and UV emission is a powerful tool to
determine the dust attenuation of light at all wavelengths (see Buat
et al$.$ 2005 and references therein). In this sense, it is worth
emphasizing that dust attenuation is the most vexing problem that one
has to face when analyzing the observational properties of composite
stellar populations and galaxies.

Finally, the observation of nearby galaxies in the UV is fundamental
if we are to understand the evolution of galaxies from the
high-redshift Universe (where their properties are commonly derived
from rest-frame UV observations) to the present.

There have been many attempts in the past to address some of these
issues. Sullivan et al$.$ (2000, 2001, 2004) studied the star
formation histories in a relatively large and complete sample of
UV-selected local galaxies, from which Treyer et al$.$ (1998) derived
the SFR density of the local Universe. The nature of the UV upturn in
elliptical galaxies has been widely studied by several groups,
including O'Connell (1999), Brown et al$.$ (2000), Deharveng, Boselli,
\& Donas (2002). Studies on the dust attenuation in galaxies based on
either photometric or spectroscopic UV studies are numerous, including
Calzetti et al$.$ (1994), Heckmann et al$.$ (1995), Meurer et al$.$
(1995, 1999), Buat \& Xu (1996), Gordon et al$.$ (2000, 2003), Buat et
al$.$ (2002), Roussel et al$.$ (2005). The analysis of the UV
morphology of nearby galaxies as a local benchmark for studies in the
optical at high redshift have been also carried out by several
authors, including Kuchinski et al$.$ (2000, 2001), Marcum et al$.$
(2001), Windhorst et al$.$ (2002), Lauger, Burgarella, \& Buat (2005).

However, the results of some of these studies were not conclusive
mainly due to the small size of the samples used, which were not
representative of the overall population of galaxies in the local
Universe. This is particularly true for studies on the dust
attenuation in star-forming galaxies and on the rest-frame UV
morphology in nearby galaxies. In the case of the UV-upturn studies in
early-type galaxies this limitation adds to the lack of spatial
resolution and depth of previous UV data and, in some cases, to the
availability of UV data in only one band, which leads to a loss of
sensitivity to the strength of the UV upturn, best traced by the
FUV$-$NUV color (see Gil de Paz et al$.$ 2005 and references therein).

The availability of deep UV observations with moderately-good spatial
resolution for large numbers of well-known nearby galaxies is now
possible thanks to the launch of the Galaxy Evolution Explorer (GALEX)
on April 28th 2003. The compilation of GALEX UV data carried out as
part of this paper will allow us (and other researchers making future
use of this dataset) to provide fundamental clues for solving some of
the still many open questions regarding the UV properties of galaxies
in the local Universe. In particular, we will show how the strength of
the UV upturn is function of the stellar mass of the galaxy, with more
massive elliptical galaxies showing stronger UV upturns. We will also
demonstrate that in a sample like ours, which adequately represents
the bulk of the galaxy population in the local Universe, the slope of
the UV continuum is well-correlated (although with a significant
dispersion) with the infrared-to-UV ratio and, therefore, with the UV
extinction, and that the (FUV$-$$K$) color provides and excellent
segregation between early-type (ellipticals and lenticulars) and
late-type (spirals and irregulars) galaxies.

In this ``The GALEX Ultraviolet Atlas of Nearby Galaxies'' we present
surface photometry in the two GALEX ultraviolet (FUV \& NUV) bands,
providing integrated photometry and structural parameters for a total
of 1034 nearby galaxies, including extensively-studied objects like
M31, M32, M~33, M~51, M~81, M~82, M~83, M~87, M~101, etc. We compare
the UV properties of this sample with corollary data in the optical,
NIR, and FIR, available for the majority of the galaxies in the
Atlas. This comparison allows us to obtain insight into fundamental
correlations such as the `red sequence' found in the color-magnitude
diagram of ellipticals and lenticulars, and a better definition of the
IRX-$\beta$ relation in normal star-forming galaxies.

In Section~\ref{sample} we extensively describe the sample of
galaxies. Section~\ref{observations} provides a summary of the GALEX
observations. The analysis and results are given in
Sections~\ref{analysis} \& \ref{results}, respectively. The
conclusions are summarized in Section~\ref{conclusions}.

\section{Sample}
\label{sample}

\subsection{Selection}
\label{sample.selection}
The sample of objects in this Atlas includes galaxies in the GALEX
Nearby Galaxies Survey (NGS) (Gil de Paz et al$.$ 2004; Bianchi et
al$.$ 2003a, 2003b) plus galaxies serendipitously found in NGS fields
or in fields with similar or greater depth obtained as part of other
GALEX imaging surveys that have optical diameters at the
$\mu_B$=25\,mag\,arcsec$^{-2}$ isophote larger than 1\,arcmin
according to the Third Reference Catalog of Bright Galaxies (RC3
hereafter; de Vaucouleurs et al$.$ 1991). Note also that 1\,arcmin is
the apparent diameter for which the RC3 catalog is reasonably complete
(Harold G. Corwin, private communication).

As mentioned above the answers to some of the most fundamental open
questions on galaxy evolution in general and on the UV properties of
galaxies in particular are largely dependent on the (sometimes very
large) corrections for dust extinction that must be applied. With this
in mind the NGS survey was constructed to optimally sample the UV as
provided by GALEX and the FIR (where most of the UV light absorbed by
dust is re-emitted) as seen by the Spitzer Space Telescope that would
give us a bolometric view of galaxies in the Local Universe. Thus, we
began building the NGS sample using Spitzer's Reserved Observations
Catalog (ROC v2.0), which guarantees that both UV and infrared data
will be eventually made publicly available to the community for all
these galaxies. This includes the {\it Spitzer Infrared Nearby
Galaxies Survey} legacy project (SINGS; Kennicutt et al$.$ 2003), but
also data from Guaranteed Time Observations (GTO) programs like {\it
The Mid-IR Hubble Atlas of Galaxies} (PI: G. Fazio), {\it Starburst
Activity in Nearby Galaxies} (PI: G. Rieke), {\it Probing a sample of
Interacting and Ultra-luminous Galaxies} (PI: G. Fazio), etc. The
total number of (targeted) NGS galaxies represents approximately one
fourth of the total sample of 1034 galaxies in the present Atlas. The
vast majority of the UV images of the galaxies observed as part of NGS
have exposure times of 1 GALEX orbit or more ($\sim$1700\,s). See
section~\ref{observations} for a detailed description of the GALEX
spacecraft and instrument.

In order to cover a wider range of physical properties (see
Section~\ref{sample.comparison}) and taking advantage of the large
field-of-view of the GALEX instrument (1.2\,degrees in diameter) we
added to the Atlas sample all galaxies in the RC3 catalog with D25
diameter larger than 1\,arcmin that were serendipitously observed
within NGS fields and/or within other GALEX surveys of similar or
greater depth that were available to the GALEX team, namely the Medium
Deep, Deep, and Ultra-deep Imaging Surveys (MIS, DIS, and UIS,
respectively). We also added galaxies that were targeted by GALEX
because they had been observed by previous UV missions like UIT or
FOCA (as an additional calibration test for GALEX) and galaxies from
dedicated observations of the central 12\,deg$^{2}$ of the Virgo
cluster (Boselli et al$.$ 2005).

Of the original 1136 galaxies compiled 26 were found in images that
have failed the quality assurance (QA) test of the image aspect
solution that is included as part of the standard GALEX pipeline. A
total of 55 additional galaxies were excluded either because were
observed in regions of high background, high Galactic extinction,
during very short orbits and/or they showed extremely low surface
brightness in the UV. After excluding these galaxies and those objects
with no published redshift (21) we ended up with a total of 1034
galaxies; 893 of them having both FUV and NUV high-quality
observations.

\subsection{Positions, sizes, morphological types, and distances}
\label{sample.parameters}
In Table~1 we give some basic parameters for the galaxies included in
this Atlas, including their positions, sizes, distances, Galactic
color excesses [E($B-V$)], and morphological and spectroscopic types.

The coordinates shown (columns 2 and 3) correspond to those given by
NED, which are known to be more accurate than those in the RC3 catalog
from which a significant fraction of these galaxies were selected
(Harold G. Corwin, private communication). A few objects (29) had NED
coordinates that were clearly offset from the position of the galaxy
both in the UV and the DSS images. In these cases, the new, correct
positions were determined by eye after inspecting the corresponding UV
images and the Table~1 corrected accordingly. Finally, for two of the
objects in the Atlas (UGC~08650 and UGC~11859) a missing/incomplete
World Coordinate System (WCS) solution was recomputed using the
positions of nearby stars in the USNO-B catalog (Monet et al$.$
2003). Sizes are the major (column 4) and minor (column 5) axis
diameters at the $B$-band 25\,mag\,arcsec$^{-2}$ isophote (computed
from the corresponding D25 and R25 values in the RC3 catalog). For
those few objects in our sample not included in the RC3 catalog we
used the major and minor axis diameters available in NED. Position
angles (PA; column 6) missing or incorrect in the RC3 catalog were
determined by eye for a total of 90 of the galaxies in the sample
after inspecting the corresponding UV images. Note that the PA is
undefined in those galaxies for which the D25 isophote is
approximately circular according to the RC3.

The distances (columns 7, 8, and 9) to objects with heliocentric
recession velocities larger than 500\,km\,s$^{-1}$ were determined
using a Virgo-infall corrected radial velocity and a Hubble constant
of 70\,km\,s$^{-1}$\,Mpc$^{-1}$. The correction from heliocentric to
Virgo-infall corrected velocity was performed in the same way as in
the LEDA database (see also Yahil et al$.$ 1977; Theureau et al$.$
1998; Sandage \& Tammann et al$.$ 1990). The distances to a total of
801 galaxies were computed in this way. Galaxies with radial
velocities less than 500\,km\,s$^{-1}$ had distances computed from a
variety of methods, including (in approximate order of preference) the
period-luminosity (PL) relation of Cepheids, measurement of the
$I$-band magnitude of the Tip of the Red Giant Branch (TRGB), the
proper motion of masers, Surface-Brightness Fluctuations (SBF), the
Globular Cluster luminosity function, the Tully-Fisher relationship,
or the brightest stars method (see Table~1 for individual references
on each galaxy). For some galaxies believed to be members of
interacting pairs, groups or clusters we adopted the distance to the
corresponding pair, group or cluster. Examples of this are NGC~1510
and NGC~1512, NGC~5713 and NGC~5719 as members of interacting pairs;
NGC~1546, NGC~1549 and NGC~1553 as part of the Dorado group;
ESO~059$-$G007 and ESO~059$-$G010 as likely members of the NGC~2442
group; NGC~1316, NGC~1317, NGC~1381, NGC~1387, NGC~1399, etc., all
members of the Fornax Cluster. For a total of 104 galaxies in the
Virgo Cluster area the distances were either derived by adopting a
three-dimensional structure of the Virgo Cluster and subdivision into
clouds, very similar to that of Gavazzi et al$.$ (1999), or directly
from the GOLDMINE database (Gavazzi et al$.$ 2003). In the case of the
SINGS galaxies we adopted the distances given by Kennicutt et al$.$
(2003).

Galactic color excesses E($B-V$) (column 10) are those available
through NED, which correspond to those given by Schlegel et al$.$
(1998).

Regarding the morphological types (columns 11 and 12), we have adopted
those given in the RC3 catalog. Galaxies were broadly binned as
elliptical/lenticular galaxies when their morphological type, T, was
less than $-$0.5, spirals if T was between $-$0.5 and 9.5, and
irregulars when T exceeded 9.5 (see de Vaucouleurs et al$.$ 1991). For
the Blue Compact Dwarf galaxies NGC~1705, NGC~2537, NGC~3125,
NGC~4344, and NGC~4861 (see Gil de Paz et al$.$ 2003, 2005),
originally classified in the RC3 catalog as ellipticals or spirals, we
changed their morphological type to Compact Irregulars
(T=11). Regarding the morphological classification of mergers we
should mention that depending on (1) the stage of evolution of the
particular merger, (2) distance and resolution of the images from
which the system was classified, and (3) the criterion of the person
classifying the object, mergers might appear classified as (1) two
galaxies each with its own morphological type (e.g$.$ NGC~4038/4039),
(2) one single peculiar (NGC~0520) or even spiral (NGC~6052) object,
or (3) they might lack any morphological classification
(Mrk~8). Finally, spectroscopic types (column 13) were taken from NED.

In Figure~1a (top panel) we show the distribution of galaxies in the
Atlas as a function of the $B$-band apparent magnitude for different
morphological types (dark gray histogram for ellipticals, gray for
spirals, light gray for irregulars) and for the total sample
(solid-lined histogram). Figure~1b (top panel) shows the distribution
of distances. The shape of this latter distribution appears to be the
result of combining the RC3 redshift distribution (broken-line
histogram) with a large number of very nearby galaxies with distances
closer than 50\,Mpc included in this Atlas as part of NGS. It is worth
noting here that because our sample is effectively limited in
magnitude and (to a lesser extent) in diameter we might be missing a
fraction of faint, low-luminosity, low-surface brightness galaxies
compared with what we would find in a volume-limited sample of the
local Universe. In this sense, objects such as the Antlia Dwarf or the
dwarf spheroidal satellites around the Milky Way might be common in
the field but they would certainly be underrepresented in either a
magnitude or a diameter-limited sample. This limitation should be kept
in mind when comparing our results with those obtained from the
analysis of a volume-limited sample, such as the 11 HUGS sample (Funes
et al$.$ 2005) or the Virgo-cluster sample analyzed by Boselli et
al$.$ (2005).

\section{GALEX observations}
\label{observations}
GALEX is a NASA small explorer class mission that orbits the Earth at
an altitude of approximately 700\,km. The single instrument onboard
consists of a 50-cm-aperture Ritchey-Chr\'{e}tien telescope equipped
with a dichroic beam splitter that allows simultaneous observation in
two separate bands, FUV and NUV, within a circular field of view of
1.2\,degrees in diameter. The dichroic also acts as a field-aberration
corrector. The UV light is detected using microchannel plates with
crossed delay-line anodes. The effective wavelength of the two GALEX
bands are 1516 and 2267\,\AA, and their full-width at half-maximum are
269 and 616\,\AA, respectively for the FUV and NUV channels. The
observations are carried out only at night-time with a typical total
usable time per orbit of $\sim$1700\,s. For a more detailed
description of the spacecraft and the intrument the reader is refered
to Martin et al$.$ (2005) and Morrissey et al$.$ (2005).

The observations for this Atlas were carried out by the GALEX
satellite between 7 June 2003 and 29 April 2005. The typical exposure
time per field was one orbit (specific exposure times are given in
Table~2). During periods of intense solar activity, including the
historical solar storm occurred on October-November 2003, or during
sporadic overcurrent events the FUV detector was turned off to avoid
any damage of the electronics. Although in some occasions the NUV
detector had also to be turned off, in most of the cases the scheduled
observations were still carried out through the NUV channel. Due to
this, a total of 141 galaxies in this Atlas (14 per cent of the
sample) were observed only in the NUV band. These galaxies are
identified as NUV-only in Table~2. Nineteen galaxies were found to be
too faint in our FUV imaging data and were analyzed as NUV-only
targets as well. Table~2 also provides information regarding the FUV
and NUV background for each of the fields along with the mean standard
deviation of the sky and the standard deviation of the mean value of
the sky accross different regions in the field (see
Section~\ref{analysis.profiles} for details).

Using the GALEX pipeline the photon lists generated by the detectors
for each of the bands were processed to produce the corresponding
intensity maps in counts per second. The final output products of the
pipeline also include a high-resolution response map, which is the
product of the effective exposure time by the flat field at a given
position and that was used to estimate the photon noise in our
images. Note that the images taken as part of the mosaics of M~31 and
of the center of the Virgo cluster were generated using a slightly
modified version of the pipeline that, nevertheless, preserve both the
image quality and absolute flux calibration generated by the standard
GALEX pipeline.

The point spread function (PSF) of the images was found to vary as a
function of the count rate with bright point sources usually leading
to a wider PSF than faint sources and as a function of the position of
the image. For the average count rates usually obtained from nearby
galaxies and for objects located within the central 0.5\,degrees of
the GALEX field the PSF full-width at half-maximum (FWHM) is in the
range 4.0-4.5\,arcsec and 5.0-5.5\,arcsec, respectively for the FUV
and NUV bands.

\subsection{Comparison with a magnitude-limited sample}
\label{sample.comparison}

Because of the rather arbitrary criteria involved in selecting the
objects in this Atlas, a comparison of the properties of these
galaxies with the overall population of galaxies in the Local Universe
is in order if the conclusions derived from this work are to be
applied beyond the limits of this Atlas. We have therefore compared
the distribution of properties of the galaxies in the Atlas with those
included in the Nearby Field Galaxy Survey (NFGS) of Jansen et al$.$
(2000). The NFGS is composed of a total of 196 galaxies that were
selected as a representative sub-sample of the magnitude-limited CfA
survey (Huchra et al$.$ 1983).

The $B$-band magnitude distributions obtained for both the GALEX Atlas
and the NFGS (see Figure~1a) are a direct consequence of the limits of
the surveys from which they are derived. That is, while the faint-end
of the distribution for the GALEX Atlas is a result of the effective
completeness limit of $\sim$15.5\,mag inherent in the RC3 (note again
that the completeness of the RC3 is also limited to objects larger
than 1\,arcmin); galaxies in the NFGS show a sharp cutoff in their
apparent magnitudes at the limit of the CfA survey,
$B$$\sim$14.5\,mag.

In Figure~1b we compare the distribution of distances of both samples
and for the whole RC3 catalog. Both distributions are similar, with
most of the galaxies found at distances closer than 100\,Mpc, with a
relatively long tail extending to distances up to 200\,Mpc and
somewhat beyond. The peak in the NFGS distance distribution is
intermediate between the local peak in the Atlas sample associated
with NGS and the more distant peak of the (slightly deeper in
apparent magnitude) RC3 catalog.

Figure~1c shows the comparison between the D25 major-axis diameter of
our sample and the NFGS. The major-axis diameters of galaxies in the
NFGS were obtained using the UGC catalog major-axis diameters and the
morphological-type-dependent transformation coefficients given by the
Table 6 of the RC3 catalog (de Vaucouleurs et al$.$ 1991). We find
that a total of 52 galaxies in the NFGS ($\sim$27\%) are smaller than
1\,arcmin in D25 major-axis diameter and would be missed by the size
limit imposed to the serendipitous part of the Atlas sample. Of these
galaxies, approximately half are elliptical/lenticular and half spiral
galaxies.

Figure~1 demonstrates that the GALEX Atlas and the NFGS, at least in
terms of their apparent magnitudes and redshift distributions, are
sampling the same volume of the Universe and represent a similar
population of galaxies. It is therefore fair to now carry out a more
detail comparison between the intrinsic properties of the galaxies in
these two samples, including their luminosities, colors, and SFR. At
this point it is also worth noting that the relative numbers of
ellipticals/spirals/irregulars (22\%, 62\%, 8\%)\footnote{Note that
about 9\% of the galaxies in the Atlas do not have morphological types
available in the RC3.} in the GALEX Atlas are similar to those found
in the field by the NFGS (28\%, 65\%, 7\%). 

In Figure~2a we compare the distribution in $B$-band absolute
magnitude of both the GALEX Atlas and the NFGS samples. The
distribution of both ellipticals/lenticulars and irregulars is pretty
similar between both samples. However, in the case of the spirals,
although the range of properties covered is also similar, we find a
moderate excess of intrinsically bright spirals
($-$21$<$M$_B$$<$$-$20) and a small paucity of low-luminosity spirals
($-$19$<$M$_B$$<$$-$18) compared with the field for the same number of
ellipticals/lenticulars and spirals outside these luminosity
bins. Note that thanks to the large number of objects in our sample
there are still more than twice more low-luminosity spirals
(M$_B$$>$$-$19\,mag) in this Atlas than in the NFGS sample. Similar
behavior is seen when the ($U-B$) colors of both samples are compared
(see Figure~2b). In this case the Atlas sample shows a slight paucity
of relatively blue (and probably also faint) spirals. Finally, we use
the 60\,$\mu$m and 100\,$\mu$m IRAS fluxes to compute the FIR
luminosity using the recipe of Lonsdale et al$.$ (1985). The
comparison of the FIR luminosities derived for each sample shows that
they both cover the same range of properties with a comparable
distribution except for the slight excess (paucity) of high (low) FIR
luminosity spirals in the Atlas sample (see Figure~2c).

The origin of this small difference in the luminosity distribution of
our sample and that of the NFGS might due in part to the size-limit of
1\,arcmin in D25 major-axis diameter imposed to the serendipitous part
of the Atlas sample. However, since only 23\% of the spirals in the
NFGS are smaller than 1\,arcmin this effect only accounts for part of
the problem. The other reason for this difference in luminosity is
probably intrinsic to the rather heterogeneous selection criteria in
the original GALEX NGS, which basically includes all galaxies that
were in the Spizter ROC at the time the GALEX surveys were planned. In
the GALEX NGS we can find targets from many different Spitzer programs
which are in many cases biased to large, bright, nearby galaxies with
expectedly bright infra-red emission (i.e$.$ bright, nearby spirals).

\section{Analysis}
\label{analysis}

\subsection{Color Images}
\label{analysis.maps}

The left panels of Figure~3 show false-color RGB maps of the galaxies
in our sample. The images used are `asinh' scaling versions (Lupton et
al$.$ 2004) of the 2-pixel-smoothed FUV image (blue), the original NUV
image (red), and a linear combination of the two (green). The
coefficients used to obtain the green-channel image are 0.2 and 0.8,
respectively for the smoothed FUV and original NUV images. For those
galaxies with NUV-only data we give the asinh-scaled NUV image. Shown
in green is the RC3 D25 ellipse, which was originally derived from
$B$-band photometry.

In most of the cases (891 galaxies) the maps shown correspond to a
region 1.5 times the D25 major axis diameter in size. In those cases
where the UV emission is comparatively more extended than the optical
light this factor was increased up to 5$\times$D25 for the most
extreme cases. The size of the horizontal tick mark plotted at the
bottom of each map corresponds to 2\,kpc at the distance of the
galaxy, except for the case of the Phoenix Dwarf (ESO~245$-$G007)
where it represents a physical size of 0.2\,kpc. For comparison
purposes the central panels of Figure~3 show DSS-1 images for the same
field of view.

\subsection{Surface brightness and color profiles}
\label{analysis.profiles}

Using the central position, ellipticity (derived from the
corresponding axial ratio) and position angle of the D25 ellipse given
in Table~1 we compute the mean surface brightness within elliptical
annuli of fixed center and position angle increasing from 6\,arcsec in
major-axis radius to at least 1.5 times the D25 radius. The outermost
point where the surface photometry was computed corresponds to the
size of the postage stamps shown in Figure~3 and which in turn depends
on the extension of the UV emission for each individual galaxy. In
some cases a few of the outermost points had to be removed from the
profiles shown in Figure~3 because either the mean flux in the
isophote was below the level of the sky of the photometry errors were
extremely large (see below). Point sources with colors redder
than (FUV$-$NUV)=1 were automatically identified as foreground stars
and masked in the GALEX images.  These masks were then visually
inspected in order to (1) include blue foreground stars and, in a very few
cases, to (2) exclude from this automatically-generated mask the
nuclei of some galaxies that had been misclassified as foreground
stars.

In order to determine the errors in the surface photometry we used the
expressions and methodology described in Gil de Paz \& Madore
(2005). Errors in the inner parts of the profiles are commonly
dominated by photon noise, while in the very outer parts the
background subtraction uncertainties dominate. It is important to note
here that because the background in these images is very low (see
Table~2), especially in the FUV channel, the statistics of the
background are highly Poissonian, therefore the common estimators, the
mean, median and mode can be very different. In particular, the mode
in shallow FUV images can be zero. Thus, since our surface photometry
uses the mean as the measure of the flux within each isophote we
consistently use the mean of the sky as a consistent estimate of the
background, after all Sextractor-detected sources are carefully masked
(Bertin \& Arnouts 1996). We further masked all pixels in a 5$\times$5
pixel box around each detected-source pixel so as to avoid possible
contamination from the light in the extended wings of the sources.

The background was computed as the mean of the sky value of a total of
90 different regions of 4000 pixels each located around the source and
arranged in two concentric elliptical patterns at a distance never
closer than 1.5 times the size of the corresponding D25 ellipse. From
a comparison of the mean of the standard deviation within each sky
region with the standard deviation of the mean of each region we also
determine the impact of low-frequency variations in the background
(due, for example, to flat-fielding errors) on the total error budget
(see Gil de Paz \& Madore 2005 for details).

In the right panels of Figure~3 we show the FUV and NUV
surface-brightness (bottom) and (FUV$-$NUV) color profiles (top) each
corrected for Galactic extinction in units of AB magnitudes per square
arcsec along with the corresponding 1\,$\sigma$ errors. Here we have
adopted the Galactic color excesses given by Schlegel et al$.$ (1998)
and the parametrization of the Galactic extincion law given by
Cardelli et al$.$ (1989) for a total-to-selective extinction ratio of
$R_V$=3.1. The conversion factors are
A$_{\mathrm{FUV}}$=7.9$\times$E($B-V$) and
A$_{\mathrm{NUV}}$=8.0$\times$E($B-V$). The FUV surface-brightness
profile shown in Figure~3 is given in blue, the NUV profile in red,
and the (FUV$-$NUV) color profile is in green. The profiles are
plotted against the equivalent radius ($\sqrt{a\times b}$) of the
corresponding ellipse (expressed both in arcsec and kiloparsecs). The
white error bar at the top-right corner of each diagram represents the
$\pm$1$\sigma$ uncertainty on the GALEX zero points, which is
estimated to be $\pm$0.15\,mag for both the FUV and NUV channels. Note
that all panels are scaled to the same range in surface brightness and
color. Hereafter when we refer to (FUV$-$NUV) it will be the color
corrected for Galactic extinction.

In all cases, except the Antlia Dwarf, the outermost point represented
in these plots corresponds to the position beyond which either the
intensity in the image falls below the level of the sky, or the error
in the surface photometry for the NUV band is larger than 0.8\,mag,
whichever happens first. In the case of the Antlia Dwarf galaxy we
limited the radial range to that where the contamination from a
diffuse Galactic cirrus, located near the position of the galaxy, was
still found to be negligible.

\subsection{Asymptotic magnitudes, colors, and structural parameters}
\label{analysis.asymptotic}

Using the surface brightness profiles derived and the area of each
elliptical annulus we obtained the asymptotic magnitudes by
extrapolating the growth curve to infinity. We first computed the
accumulated flux and the gradient in the accumulated flux (i.e$.$ the
slope of the growth curve) at each radius and perform an
error-weighted linear fit to the accumulated flux versus slope of the
growth curve plot. After an appropiate radial range was chosen we took
the value of the $y$-intercept of this fit as the asymptotic magnitude
of the galaxy. This technique is described in detail in Cair\'os et
al$.$ (2001). These authors also tested the stability of this method
against the choice of radial range used for the fit and verified its
reliability by comparing their results with those obtained using
alternative extrapolation techniques. Note that obtaining growth curves and 
deriving the corresponding asymptotic magnitudes of galaxies at 
UV wavelengths have been already 
done in the past (see Rifatto, Longo, \& Capaccioli 1995).

Asymptotic magnitudes and colors along with their corresponding errors
are shown in one of the corners of the panels on the right of
Figure~3. These errors are composed of a term derived from the
error-weighted fit of the growth curve plus a term (in parentheses)
due exclusively to uncertaintines in the GALEX FUV and NUV zero points
($\pm$0.15\,mag). In Table~3 we give the asymptotic AB magnitudes in
both the FUV and NUV bands along with the corresponding asymptotic
(FUV$-$NUV) colors. The asymptotic luminosities (in Watts) and the
aperture magnitudes and colors inside the D25 elliptical aperture are
also provided in this table. The mean differences obtained between the
asymptotic magnitudes and the D25 aperture magnitudes are
$-$0.19$\pm$0.20\,mag and $-$0.23$\pm$0.20, respectively for the FUV
and NUV, with the asymptotic magnitudes being brighter. The errors
quoted in Table~3 correspond to the error associated with the fit to
the growth curve alone.

From the growth curve obtained we also computed the effective radius
as the equivalent radius at which the accumulated flux was equal to
the asymptotic magnitude plus 0.7526\,mag [2.5\,$\log(2)$]. In a
similar way we derived the radii containing the 20, 25, 75, and 80~per
cent of the light ($r_{20}$, $r_{25}$, $r_{75}$, and $r_{80}$,
respectively) that were used to compute the concentration indices C31
(de Vaucouleurs 1977) and C42 (Kent 1985) in both UV bands. These
indices are defined as
\begin{eqnarray}
\mathrm{C31} = {{r_{75}} \over {r_{25}}}\\
\mathrm{C42} = 5 \log{ \left( {{r_{80}} \over {r_{20}}} \right) }
\end{eqnarray}

\subsection{Morphological classification of the UV profiles}
\label{analysis.morphology}

We have visually classified the UV surface brightness profiles shown
in Figure~3 according to their shape. Since most of the profiles
(especially in spiral and irregular galaxies) show two distinct
regions, our classification scheme uses two letters: the first letter
describing the shape of the outer profile and the second one
describing the shape of the inner region. In a few cases where we find
an excess or depression associated with the nucleus of the galaxy we
add a final suffix {\bf n} ({\it nucleated}) or {\bf h} ({\it hole}),
respectively, to the corresponding morphological class. Also in
galaxies showing obvious extended UV emission (XUV; Thilker et al$.$
2005; Gil de Paz et al$.$ 2005) the morphological class is preceded by
the letter {\bf x} ({\it eXtended}). The classes assigned are given in
column~16 of Table~3. Those galaxies that are barely resolved by our
GALEX observations have no such classes assigned. The codes used for
the morphological classification of the outer region are: {\bf E} for
exponential, {\bf V} for a de Vaucouleurs profile, or {\bf ?} if there
are not enough points in the outer profile to determine which of the
two previous laws works best. In the case of the inner profile we use:
{\bf E} or {\bf V} if the profile is a smooth continuation of the
corresponding outer profile or if there is a transition from a {\bf E}
profile in the outer parts to {\bf V} in the inner regions, {\bf F}
for a flattening of the profile toward the inner regions, {\bf D} for
a profile falling in brightness toward the center, and finally {\bf R}
for a profile moderately rising in brightness over what it would be
expected from and inward extrapolation of the outer profile law. The
letter describing the shape of the inner profile appears in lower case
if the radial extension of the inner profile is significantly smaller
than that of outer profile. In this scheme galaxies with pure de
Vaucouleurs (exponential) profiles would be classified as VV (EE)
type. Note that the majority of the galaxies in this Atlas are
extracted from GALEX fields of similar depth ($\sim$1\,orbit) that
were obtained as part of the Nearby Galaxies Survey (NGS) or the
Medium-deep Imaging Survey (MIS). In the majority of the cases the
same classification does apply to both the FUV and NUV profiles. In
the few cases where the profiles differ enough to be placed in
different classes we give first the FUV and then the NUV morphological
class separated by a comma (e.g$.$ NGC1055, NGC1386, NGC1546)

\subsection{Corollary data}
\label{analysis.corollary}

In order to compare the UV properties of the galaxies in this Atlas
with those known from previous multi-wavelength surveys we have
compiled a large amount of corollary data on this sample (see
Table~4). Of the 1034 galaxies in the Atlas a total of 871 (84\%) have
asymptotic $B$-band photometry available in the RC3 catalog. We
primarily used the $B_T$ magnitude and only when $B_T$ was unavailable
we made use of the m$_B$ magnitude instead. A total of 318 (393)
galaxies also have asymptotic $U$ ($V$) magnitudes published in the
RC3.

In addition we have also compiled integrated $JHK$ magnitudes from
2MASS. In the first instance we adopted the $JHK_{\mathrm{tot}}$
magnitudes from the 2MASS Large Galaxy Atlas (LGA) of Jarrett et al$.$
(2003). For those objects not in the 2MASS LGA we used the total
$JHK_{\mathrm{total}}$ magnitudes given in the Final Release of the
2MASS Extended Source Catalog (XSC). A total of 853 galaxies in the
Atlas had $K$-band data available.

The optical and near-infrared magnitudes given in Table~4 are observed
values. The corresponding Galactic extinction-corrected magnitudes were
derived using the color excesses given in Table~1 and the extinction law
of Cardelli et al$.$ (1989) for $R_V$=3.1.

Finally, we compiled IRAS photometry using data from (in order of
priority) Rice et al$.$ (1988), Knapp (1994, private communication),
the IRAS Point Source Catalog (PSC), and Moshir et al$.$ (1990). A
total of 459 galaxies had IRAS detections at both 60 and
100\,micron. These two bands are required in order to estimate the
total infrared emission of the galaxy and from it the total energy
budget by means of its comparison with the UV flux (see e.g$.$ Dale et
al$.$ 2001).

\section{Results}
\label{results}

\subsection{Global statistical properties}
\label{results.statistical}

In Figure~4 we show the frequency histograms of the asymptotic FUV and
NUV AB magnitudes, FUV luminosity and (FUV$-$NUV) (both asymptotic and
at the D25 aperture) color. Heckman et al$.$ (2005) have recently
shown that galaxies with FUV luminosities brighter than
2$\times$10$^{10}$\,L$_{\odot}$ (7.6$\times$10$^{36}$\,W or
M$_{\mathrm{FUV}}$=$-$19.87) (also known as ultraviolet-luminous
galaxies or UVLGs) are extremely rare in our Local Universe. Their
comoving space density is only $\sim$10$^{-5}$\,Mpc$^{-3}$, i.e$.$
several hundred times lower than that of their $z$=3 counterparts, the
Lyman Break Galaxies (LBG). Indeed, only four galaxies in the Atlas
(see Figure~4c) would be classified as UVLGs: two AGN, NGC~7469 and
Mrk~501, and two actively star-forming interacting systems, the
Cartwheel (see e.g$.$ Amram et al$.$ 1998) and UGC~06697 (Gavazzi et
al$.$ 2001).

The color distribution of Figure~4d shows a pronounced peak at
(FUV$-$NUV)$\simeq$0.4\,mag and a long tail extending to very red
colors. As we will show later, this red tail is, not unexpectedly,
mostly populated by elliptical galaxies of intermediate mass that show
little recent star formation activity and a weak UV-upturn (see 
Boselli et al$.$ 2005). This figure also shows the distribution of
effective radii both in arcsec (Figure~4e) and in kiloparsecs
(Figure~4f). The distribution of effective radii is very similar for
the FUV and the NUV. Due to the limited spatial resolution of the
GALEX data we only computed the effective radius of galaxies for which the semi-major axis
of the ellipse including 50~per cent of the light was larger than
6\,arcsec in radius. This fact, along with the lower limit in optical
diameter (1\,arcmin) imposed by the completeness of the RC3, results
in a paucity of compact galaxies and a relatively narrow distribution
in apparent effective radius peaking at $\sim$15\,arcsec. The
distribution in physical size (Figure~4f), on the other hand, is
significantly wider with a peak around 5-6\,kpc.

The distributions of the concentration indices C31 and C42 (Figures~4h
\& 4i, respectively) are also very narrow with the galaxies being
slightly more concentrated (i.e$.$ larger values of C31 and C42) in
the NUV than in the FUV (see Figures~4j \& 4k for a comparison 
between the value of these indices in the two bands). This is probably a
consequence of the fact that in the NUV a significant fraction of the
light in spiral galaxies still arises from within the bulge component,
while in the FUV this contribution is in many cases negligible.

\subsection{Properties by morphological type}
\label{results.morphological}

The GALEX FUV and NUV observations presented here, along with the
corresponding corollary data in the optical, NIR and FIR provides us
with an unprecedented set of multiwavelength data for a large
population of galaxies in the local Universe. One of the first
questions that can be addressed using this sample concerns the
relation between the qualitative (optical) morphology of these
galaxies and more quantitative properties, such as colors,
luminosities, total-infrared-to-UV ratios, etc. In Figure~5 we show
the colors of the galaxies as a function of the blue-light
morphological type as given by the RC3. Panels 5a \& 5b show that
although late-type spiral and irregular galaxies are somewhat bluer in
($B-V$) and ($B-K$) than ellipticals and early-type spirals, these
colors are not unique to a given type. In particular, these colors
cannot be used to unambiguously discriminate between different kinds
of spiral galaxies nor even between elliptical/lenticular galaxies and
spirals. As indicated by Roberts \& Haynes (1994), the significant
ovelap in ($B-V$) color between spiral galaxies of different types is
mostly due to true variations in the optical colors and star-formation
history of galaxies of same morphological type, not to
misclassification or observational errors. The equivalent to the
Panel~5b for late-type Virgo cluster galaxies was obtained by Boselli
et al$.$ (1997). These authors obtained a large overlap in ($B-K$)
color between different morphological types as well.

However, thanks to the extreme sensitivity of the FUV data to the
presence of very low levels of recent star formation activity, the use
of the (FUV$-K$) color turns out to be a very powerful discriminant
between quiescent elliptical and lenticular galaxies, and star-forming
spirals. In particular, an observed (FUV$-K$) color of 8.8\,mag
provides an excellent discrimination point between these two groups
(see Figure~5c). In this sense, of all the elliptical/lenticular
galaxies in the Atlas with both FUV and $K$-band data available only
23\% of them show a (FUV$-K$) color bluer than this threshold. It is
worth noting that significant a fraction of these are known to host
some residual star formation activity (e.g$.$ NGC~3265, Condon,
Cotton,
\& Broderick 2002 and NGC~0855, Wiklind, Combes, \& Henkel
1995), or are low-luminosity ellipticals with obvious star formation
activity like NGC~1510 (Marlowe, Meurer, \& Heckman 1999). Spiral and
irregular galaxies with (FUV$-K$) colors redder than this value only
represent 9\% of the total.

Although with significantly degraded discriminating capabilities
compared to the (FUV$-K$) color, the (NUV$-K$) is also well correlated
with the morpholotical type (see Figure~5d). The same can be said
about the (FUV$-$NUV) color, where a cut-off at (FUV$-$NUV)=0.9\,mag
provides a relatively clean separation of elliptical/lenticular
galaxies from spirals (Figure~5e). The fraction of
elliptical/lenticular galaxies with (FUV$-$NUV) color bluer than
0.9\,mag (and both FUV and NUV magnitudes available) is 18\% while the
percentage of spiral and irregulars redder than this value is only
12\%. Note that in this case the far-left lower corner of the diagram
may be populated both by ellipticals with residual star formation and
also by elliptical galaxies with a strong UV-upturn (Deharveng,
Boselli, \& Donas 2002 and references therein). The best linear fits
derived for the correlation of observed colors with the morphological
type for spirals and irregulars (types T$>$$-$0.5) are
\begin{eqnarray}
(\mathrm{FUV}-K) = 7.97 - 0.48 \times T\ \ \ \ ; \ \ \ \ \sigma = 1.36\,\mathrm{mag}\\ 
(\mathrm{NUV}-K) = 7.07 - 0.40 \times T\ \ \ \ ; \ \ \ \ \sigma = 1.14\,\mathrm{mag}\\ 
(\mathrm{FUV}-\mathrm{NUV}) = 0.854 - 0.066 \times T\ \ \ \ ; \ \ \ \ \sigma = 0.32\,\mathrm{mag}
\end{eqnarray}
These relations are shown in Figures~5c, 5d, \& 5e. Note that although
the r.m.s$.$ of the fit for the (FUV$-$NUV) color is smaller than for
(FUV$-K$) this is purely a consequence of the much smaller dynamic
range of the (FUV$-$NUV) color (1\,mag) compared with the (FUV$-K$)
color ($\sim$6\,mag) (see Figure~5c). The corresponding best fits in
the type T versus color diagrams are (only galaxies with types T$<$13
are considered)
\begin{eqnarray}
T = 11.2 - 1.28 \times (\mathrm{FUV}-K)\ \ \ \ ; \ \ \ \ \sigma = 2.4\ \ \ \mathrm{(in\ units\ of\ T)}\\ 
T = 12.1 - 1.62 \times (\mathrm{NUV}-K)\ \ \ \ ; \ \ \ \ \sigma = 2.5\ \ \ \mathrm{(in\ units\ of\ T)}\\ 
T = 8.4 - 8.5 \times (\mathrm{FUV}-\mathrm{NUV})\ \ \ \ ; \ \ \ \ \sigma = 3.0\ \ \ \mathrm{(in\ units\ of\ T)}
\end{eqnarray}
These fits are valid only for colors (FUV$-K$)$<$8.8\,mag,
(NUV$-K$)$<$7.9\,mag, and (FUV$-$ NUV)$<$0.9\,mag, respectively.

Finally, in Figure~5f we compare the total-infrared (TIR hereafter) to
FUV ratio with the morphological type of the galaxies in the
Atlas. The TIR flux was derived using the parameterization of the
TIR-to-FIR ratio given by Dale et al$.$ (2001), where FIR is computed
from the 60 and 100\,micron IRAS fluxes as in Lonsdale et al$.$
(1985). The flux in the FUV is expressed in units of $\nu$F$_{\nu}$
(see Buat et al$.$ 2005). In the case of spiral and irregular
galaxies, for which both the UV and infrared emission are ultimately
due to young massive stars, this ratio provides a well defined
estimator of the dust attenuation in the UV (Buat et al.$.$ 2005;
Cortese et al$.$ 2006). Given the sensitivity limits of the IRAS
catalog and the low dust content of elliptical and lenticular galaxies
the number of these galaxies detected in both the 60 and 100\,micron
IRAS bands is only 49 out of the 225 ellipticals in the
Atlas. Figure~5f shows that late-type spirals and irregulars tend to
show, on average, a lower TIR-to-FUV ratio and consequently smaller
attenuation in the UV than that derived for early-type spirals.

\subsection{Color-magnitude and color-color diagrams}
\label{results.cmd}

Although morphology is certainly related with the way galaxies form
and evolve, especially when the properties of elliptical and spiral
galaxies are compared, the luminosity and even more the mass (either
the luminous or total mass) is thought to be the main driving force of
the evolution of galaxies through the history of the Universe. In this
sense, the analysis of color-magnitude diagrams (CMD) has
traditionally provided a fundamental tool for understanting galaxy
evolution.

Figures~6a and 6b show the CMD in (FUV$-K$)-M$_K$ and
(NUV$-K$)-M$_K$. At the top of these diagrams we find the `red
sequence' populated primarily by elliptical and lenticular galaxies
(dots). In the case of the (NUV$-K$)-M$_K$ CMD the red sequence shows
a clear slope with lower luminosity galaxies showing bluer colors,
especially below M$_K$$>$$-$23\,mag. A similar behavior is seen when
optical or optical-NIR colors are used, both locally and at high
redshift (Gladders \& Yee 2005). This is commonly explained in terms
of lower metal abundances (thus bluer colors) of the stellar
populations in low mass ellipticals as compared to the more massive
(higher metallicity) systems (Gladders et al$.$ 1998 and references
therein). In the case of the (FUV$-K$)-M$_K$ CMD, on the other hand,
the distribution of the (FUV$-K$) color is rather flat over a range of
almost 7\,mag in absolute magnitude. The explanation for this
different behavior can be found in Figure~6c. Here the (FUV$-$NUV)
gets systematically redder as we move to lower luminosities. This is
opposite to what is seen in any other colors and it is probably a
consequence of a weaker UV-upturn in intermediate-mass ellipticals
than in the most luminous and massive ones (see Boselli et al$.$
2005). Note that, due to the stronger UV-upturn towards the centers of
elliptical galaxies (Ohl et al$.$ 1998; Rhee et al$.$ 2006, in
preparation), the asymptotic colors do not probably show the full
strength of the UV-upturn in the way aperture colors like those
obtained from the analysis of IUE spectra do (Burstein et al$.$ 1988).

Dwarf elliptical galaxies have $K$-band absolute magnitudes that are
typically fainter than M$_K$=$-$21\,mag. Unfortunately, not many of
these more extreme low-luminosity ellipticals are found in the
Atlas. This is mainly because dwarf ellipticals in Virgo (where most
of the studies on dE have been carried out to date) are typically
smaller than 1\,arcmin in size placing them outside the selection
limit imposed on the Atlas. Nevertheless, a recent study by Boselli et
al$.$ (2005) suggests that residual star formation might play a
leading role in the interpretation of the UV emission from dE
galaxies, which would explain their behavior in the CMD (i.e$.$
similar to the behavior seen in low mass star-forming galaxies). The
tendency for the most luminous ellipticals to show bluer (FUV$-$NUV)
colors is even more clear when the FUV-band absolute magnitude is
considered (see Figure~6e). However, if the $B$-band luminosity is
used, the (FUV$-$NUV) color seems to be independent of luminosity.

Regarding the properties of spiral (triangles) and irregular galaxies
(asterisks) in these plots we find that the majority of these galaxies
are concentrated in a `blue sequence' with high-luminosity spirals
(which also tend to be of earlier types) being redder than low-mass
spirals and irregular/compact galaxies. This is true for all the
observed (FUV$-K$), (NUV$-K$), and (FUV$-$NUV) colors (Figures~6a, 6b,
\& 6c). There are two mechanisms that may lead to the observed
behavior. First, low luminosity galaxies are known to have lower
metallicities (both in the stars and in the gas) than more luminous
ones (Salzer et al$.$ 2005 and references therein). This implies that
the amount of dust (and reddening of the colors) in low-luminosity
galaxies should be lower than in luminous ones.

The (FUV$-$K) [(NUV$-$K)] color is found to span a range of 5\,mag
[4\,mag] in spiral and irregular galaxies of different types and
luminosities with a mean value of 5.9\,mag [5.4\,mag]. The
corresponding 1-sigma of the distribution is 1.7\,mag [1.4\,mag]. On
the other hand, the dispersion in the A$_{\mathrm{FUV}}$
[A$_{\mathrm{NUV}}$] derived is only 1.0\,mag [0.8\,mag] (see
below). Since the A$_{\mathrm{FUV}}$/(A$_{\mathrm{FUV}}$$-$A$_{K}$)
[A$_{\mathrm{NUV}}$/(A$_{\mathrm{NUV}}$$-$A$_{K}$)] total-to-selective
extinction ratio is always between 1.0 and 1.1 for any attenuation law
considered, dust extinction alone is not able to explain the
dispersion in the observed (FUV$-$K) [(NUV$-$K)] color neither its
dependence on luminosity or morphological type.

It is now widely accepted that the star formation history of galaxies
depends strongly on their stellar or total mass. Low mass galaxies
show relatively flat star formation histories, while more massive
systems have shorter timescales of formation (e.g$.$ Gavazzi et al$.$
1996, 2002; Gavazzi \& Scodeggio 1996; Boselli et al$.$ 2001). By
virtue of this phenomenon, sometimes simplistically referred to as
'down-sizing' (see Cowie et al$.$ 1996), low-mass galaxies should be
on average bluer in these colors than more massive galaxies. In this
sense, we know that the typical stellar mass of a star-forming galaxy
in the local Universe is $\sim$1.3$\times$10$^{10}$\,M$_{\odot}$
(P\'erez-Gonz\'alez et al$.$ 2003; Gil de Paz et al$.$ 2000), i.e$.$
more than five times less massive than a L$^*$ galaxy in the NIR (Cole
et al$.$ 2001; Kauffmann et al$.$ 2003).

Since we have information about the TIR emission for a large fraction
of these galaxies we can compute the attenuation in the FUV and NUV
from the observed TIR-to-FUV ratio using the recipes published by Buat
et al$.$ (2005). The mean and 1-sigma FUV [NUV] attenuation of the
sample of spiral and irregular galaxies in the Atlas is
1.8$\pm$1.0\,mag [1.3$\pm$0.8\,mag]. The extinction-corrected
(FUV$-$NUV) color is plotted in Figure~6d as a function of the
$K$-band absolute magnitude. The solid (dashed) line shown in this
plot represents the best weighted (non-weighted) fit to the data
\begin{eqnarray}
(\mathrm{FUV}-\mathrm{NUV})_0 = 0.1083 + 0.00371 \times \mathrm{M}_K\ \ \ \ ; \ \ \ \ \sigma = 0.054\,\mathrm{mag}\ \ \mathrm{(weighted)}\\
(\mathrm{FUV}-\mathrm{NUV})_0 = 0.0942 + 0.00299 \times \mathrm{M}_K\ \ \ \ ; \ \ \ \ \sigma = 0.055\,\mathrm{mag}\ \ \mathrm{(non-weighted)}
\end{eqnarray}
Although there is a small tendency for the galaxies to show redder UV
colors at lower luminosities and later types, we do not exclude the
possibility that the intrinsic (FUV$-$NUV) color derived in this way
is independent of luminosity with an average value of
(FUV$-$NUV)$_0$= 0.025$\pm$0.049\,mag (i.e$.$ $\beta_{\mathrm{GLX},0}$=
-1.94$\pm$0.11; see Kong et al$.$ 2004). We should note here that the
measurements of the extinction in the FUV and NUV from which this
intrinsic (FUV$-$NUV) color is derived are not fully independent since
both are obtained by comparing the corresponding observed FUV and NUV
flux with the same total-infrared emission (Buat et al$.$
2005). Consequently, there might be some additional weak dependency of
the intrinsic (FUV$-$NUV) color with the luminosity that could be
identified by analyzing both the detailed star formation history and
dust properties (composition, geometry, temperature distribution) of
individual galaxies.

Figure~6e shows that the most luminous galaxies in the FUV are spirals
(both early- and late-type ones). In the optical (Figure~6f) and NIR
(Figure~6c), on the other hand, the bright end of the luminosity
function is populated by both elliptical and spiral galaxies. It is
also worth noting that the galaxies in the bright end of the FUV
luminosity function show a very narrow dispersion in the observed
(FUV$-$NUV) color, that results in a very similar shape for the bright
end of the FUV and the NUV local luminosity functions (Wyder et al$.$
2005).

In Figure~7a we analyze the (FUV$-$NUV)-(NUV$-B$) color-color diagram
of the galaxies in the Atlas. It is remarkable the relatively narrow
strip of this diagram where the galaxies are located. In the case of
the spiral galaxies this is due in part to the well-known degeneracy
in these colors between dust extinction and star formation history
(see e.g$.$ Gil de Paz \& Madore 2002). The ellipticals show a very
narrow range in (NUV$-B$) color but a wide range of (FUV$-$NUV)
colors, probably due to differences in the strength of the UV-upturn
from galaxy to galaxy. In the (FUV$-$NUV)-(NUV$-K$) color diagram
(Figure~7b) we find that ellipticals with redder (NUV$-K$) color tend
to show bluer (FUV$-$NUV) colors. This is again a consequence of the
weaker UV-upturn present in optically blue, intermediate-mass
ellipticals. The combination of the (FUV$-$NUV) color with either the
(NUV$-B$) or the (NUV$-K$) color clearly improves the discrimination
between elliptical/lenticular galaxies and spirals (see broken lines
in Figures~7a \& 7b). In the case of the (FUV$-$NUV)-(NUV$-B$)
color-color diagram the origin \{destination\} of the cut-off line is
[(FUV$-$NUV),(NUV$-B$)]=[2.0,2.0] \{5.0,0.0\}. For the
(FUV$-$NUV)-(NUV$-K$) color-color diagram the corresponding origin
\{destination\} of the cut-off line is [(FUV$-$NUV),(NUV$-K$)]=[5.0,1.7]
\{9.5,0.4\}.

\subsection{Dust extinction and the IRX-$\beta$ relation}
\label{results.dust}

The relation found by Heckman et al$.$ (1995) and Meurer et al$.$
(1995, 1999) between the TIR-to-FUV ratio and the slope of the UV
spectrum in starburst galaxies (IRX-$\beta$ relationship; see also
Seibert et al$.$ 2005) can be used in principle to estimate the dust
extinction in galaxies even if FIR data are not available. Some recent
works have claimed that this relationship is valid only when applied
to UV-selected starburst galaxies but not in the case of
infrared-bright objects like the luminous/ultra-luminous infrared
galaxies (LIRGs/ULIRGs; Goldader et al$.$ 2002) or even for normal
spiral or irregular galaxies (see Bell et al$.$ 2002 for results on
the LMC). In Figure~8a we compare the TIR-to-FUV ratio with the
observed (FUV$-$NUV) color, which is equivalent to the slope of the UV
continuum (see Kong et al$.$ 2004). Here we have only plotted galaxies
with observed (FUV$-$NUV) color bluer than 0.9\,mag. This criterion
guarantees that the vast majority of the objects considered are either
spiral or irregular galaxies. The dotted line represents the
IRX-$\beta$ relation given by Meurer et al$.$ (1999). This figure
demonstrates that the slope of the UV is indeed well correlated with
the TIR-to-FUV and can be used to estimate (at least in a statistical
way) the dust extinction in nearby galaxies. Similar results are found
by Cortese et al$.$ (2006) using a volume-limited optically-selected
sample of galaxies in nearby clusters.

The solid line in Figure~8a represents the best linear fit to the
data. The dashed line is the same but excluding objects with
luminosities below 0.1$\times$L$^*$ (M$^*_\mathrm{FUV}$=$-$18.12;
Wyder et al$.$ 2005), for which the relation begins to depart from
linearity. The results of these fits are
\begin{eqnarray}
\log(\mathrm{TIR}/\mathrm{FUV}) = -0.18 + 2.05 \times (\mathrm{FUV}-\mathrm{NUV})\ \ \ \ ;\ \ \ \ \sigma=0.36\,\mathrm{dex}\\
\log(\mathrm{TIR}/\mathrm{FUV}) = -0.15 + 2.00 \times (\mathrm{FUV}-\mathrm{NUV})\ \ \ \ ;\ \ \ \ \sigma=0.36\,\mathrm{dex}\ \ \ (\mathrm{for}\ \mathrm{L} > {{\mathrm{L}^*} \over {10}})
\end{eqnarray}
Note that our sample suffers of a small deficiency of low-luminosity
spirals. This fact might have an impact on the best-fit IRX-$\beta$
relationship derived above. Cortese et al$.$ (2006) have recently
proposed a set of recipes that can be used to estimate the TIR-to-FUV
ratio in star-forming galaxies using not only the (FUV$-$NUV) color
but other parameters such as the oxygen abundance, the luminosity, the
mean surface brightness, etc.

The majority of the objects in Figure~8a are found below the
relationship defined for starburst galaxies. It is worth noting that
objects with higher UV luminosity, some of them starburst galaxies,
seem to fall closer on average [at least in the region with
(FUV$-$NUV)$<$0.6\,mag] to Meurer et al.'s relation than lower
luminosity galaxies. According to Kong et al$.$ (2004) the offset
between normal galaxies and starbursts is primarily due to a lower
ratio of present to past-averaged SFR in normal galaxies. However, the
results obtained by Seibert et al$.$ (2005) and Cortese et al$.$
(2006) using GALEX data of nearby galaxies do not support this
idea. These recent studies suggest that this offset might be due
instead to a different geometry of the dust in normal galaxies
compared with starbursts or, alternatively, to aperture effects
present in the IUE dataset used by Meurer et al$.$ (1999).

The fact that we find such a good correlation between the TIR-to-FUV
ratio and the (FUV$-$NUV) color and that the intrinsic (FUV$-$NUV)
color seems to be rather constant for spiral and irregular galaxies
suggests that the attenuation law in the UV for these galaxies is
different from a pure Galactic extinction law. In the case of the
Milky Way the extinction law shows a bump at 2175\,\AA\ that would
result in a similar extinction in both bands,
A$_{\mathrm{FUV}}$=7.9$\times$E($B-V$) and
A$_{\mathrm{NUV}}$=8.0$\times$E($B-V$) (Bianchi et al$.$ 2005). Thus,
the observed trend in the (FUV$-$NUV) color with the TIR-to-FUV ratio
is most probably due to a different extinction law since scattering,
either for a shell or clumpy dust geometry, would result in an even
lower FUV attenuation (compared with the NUV) than that expected from
the Galactic extinction law alone (see e.g$.$ Roussel et al$.$
2005). The SMC Bar or 30~Doradus extinction laws and the attenuation
law proposed by Calzetti et al$.$ (1994) all show a weak 2175\,\AA\
feature and, especially in the case of the SMC Bar extinction law, a
relatively steep FUV rise. In this sense, despite of including
scattering, the FUV rise of the Calzetti law is apparently too modest
to reproduce the dependence between A$_{\mathrm{FUV}}$ and
A$_{\mathrm{FUV}}$$-$A$_{\mathrm{NUV}}$ followed by the majority of
the galaxies in our sample (see Figure~8b)\footnote{We have adopted
$R_V$=3.1 for the Milky Way and LMC 30~Doradus extinction laws
(Cardelli et al$.$ 1989), $R_V$=4.05 for the Calzetti law (Calzetti et
al$.$ 2000), $R_V$=2.87 for the SMC Bar law (star AzV~398; Gordon \&
Clayton 1998), and $R_V$=2.66 for the SMC Wing law (star AzV~456;
Gordon \& Clayton 1998)}. Thus, although the Calzetti law, originally
built for UV-bright starburst galaxies, still provides an adequate
approximation to the relation between A$_{\mathrm{FUV}}$ and
A$_{\mathrm{FUV}}$$-$A$_{\mathrm{NUV}}$ for galaxies with UV
luminosities above $L^{*}$, an attenuation law based on the SMC-Bar
extinction law is favored for the bulk of the galaxies in this Atlas.

We cannot exclude, however, that the FUV emission might be arising
from young stars more deeply embedded in their parent molecular clouds
than those responsible for the NUV emission. If that is the case, the
differential extinction between the FUV and NUV emitting sources would
lead to an artificial FUV rise in the global attenuation law even if
the extinction law is rather flat in the UV.

In this same sense, it is worth noting that here we are referring to
the attenuation law of the dust associated with the regions
responsible for the UV emission, which could be quite different from
the law we would obtain from regions dominating the emission at other
wavelengths and also different from the extinction law that would be
derived from line-of-sight absorption studies of individual stars.

\subsection{Structural properties and UV morphology}
\label{results.structural}
Concentration indices have been commonly used in the past to infer the
morphological types of barely resolved intermediate redshift galaxies
found in HST images (see e.g$.$ Abraham et al$.$ 1996). One of the
problems associated with these studies is the fact that in many cases
the concentration indices derived for the high-redshift galaxies are
measured in the rest-frame UV while the local reference samples are
usually observed in the optical (Bershady, Jangren, \& Conselice
2000). In this sense, it is important to know the structural
parameters in the UV of a sample of well-known nearby galaxies, like
the one collected for this Atlas. Concentration indices C31 and C42
are provided in Table~3. The number of objects with these indices is
small because we only computed the C31 (C42) concentration index for
those galaxies whose radius containing 25\% (20\%) of the light was
larger than 6\,arcsec. The same criterion applies to the effective
radius, where the radius containing 50\% of the light was imposed to
be larger than 6\,arcsec in order for it to be measured.

In Figure~9 we compare the concentration index C42 with the (FUV$-K$)
color. As we commented in Section~\ref{results.morphological} this
color discriminates very well between elliptical and spiral galaxies
and also between spiral galaxies of different types (see
Section~\ref{results.morphological}). This figure shows that the C42
index improves the discrimination between ellipticals (dots) and
lenticulars (open circles) and also between these and early-type
spirals (open triangles). Joe et al$.$ (2006, in preparation) have
recently carried out a more detail study of the structural properties
(including both concentration and asymmetry parameters) of nearby
galaxies in the UV using the same sample presented in this Atlas.

Regarding the morphological classification of the UV
surface-brightness profiles we first notice a large variety of
morphologies even within each of the classes defined in
Section~\ref{analysis.morphology}. This is partly a consequence of the
high sensitivity of the UV to the recent star formation which results
in the presence of structures having relatively short evolutionary
time-scales that might dominate the UV profiles but that are not as
obvious in the optical or NIR profiles. There is also the difficulty
of dealing with degeneracies between some morphological classes. In
this sense, some of the Blue Compact Dwarf galaxies in the sample
could be easily classified as having ER or EV profiles. Also, some of
the profiles inspected could be either classified as EEh or
Ed. Despite of these issues we successfully classify the profiles of
970 of the 1034 galaxy in the Atlas. Moreover, we find that most of
the galaxies (615 out of 970) have UV profiles that can be grouped in
three main classes: (1) profiles that can be reproduced entirely by a
de Vaucouleurs law (class VV), (2) pure exponential profiles (class
EE), (3) profiles with an exponential component in the outer region
and significant flattening in the inner region (EF and Ef
classes). Only 19 galaxies were classified as EV class, despite being
the dominant morphology in the optical and near-infrared profiles of
spiral galaxies.

This paucity of EV profiles seems to be due, at least in the case of
late type spirals, to the fact that even in the central regions the
bulge is much fainter than the disk, which results in these galaxias
being classified as having type EE or EF/Ef profiles (e.g$.$ NGC~0628,
M~33, NGC~1042, NGC~2403). In early-type spirals, like the Sb galaxies
NGC~0986, M~31, M~81, M~95, the bulge is dominant only in the nucleus
of the galaxy where is also commonly found associated with a
flattening or decrease in the surface brightness of the disk toward
the center. Because of the small spatial extension of these bulges in
the UV surface brightness profiles Sb galaxies get usually classified
as EFn, VFn, or EDn. Only lenticulars (e.g$.$ NGC~1387, NGC~1546,
NGC~4310, NGC~4477, NGC~6945, NGC~7252, M~86), intermediate S0/a
(NGC~2681, NGC~3816, NGC~3885, IC~0796), or very early-type spirals
like the Sa galaxies NGC~1022, NGC~2798, NGC~4314, or NGC~4491, are
sometimes best classified as having EV-type UV profiles. This is true
for both UV bands although it is more frequent in the case of the NUV
profile.

In Figure~10 we plot the distribution of galaxies classified within
each of these groups: de Vaucouleurs profiles ({\bf v}), pure
exponential profiles ({\bf e}), and flattened exponential profiles
({\bf f}); in the (FUV$-K$) versus morphological type diagram. Again,
the morphological types used are those published in the RC3. In the
light of this figure it is fair to say that the majority of the
elliptical galaxies in the Atlas follow a de Vaucouleurs profile in
the UV, like is the case of the optical and NIR profiles of luminous
elliptical galaxies. Note that because of our selection limits a small
number of dwarf elliptical galaxies (which commonly show exponential
light profiles in the optical) is expected to be found in this
Atlas. A few {\bf v}-type galaxies classified morphologically as late
T-type objects are found to be well-known Blue Compact Dwarf (BCD)
galaxies: NGC~1569, NGC~3125, NGC~5253, NGC~6789, UGC~05720 (Haro~2).
See Doublier et al$.$ (1997, 1999) for some other examples of BCD
galaxies with R$^{1/4}$ profiles in the optical.

Regarding the distribution of the other two types of profiles we point
out that while galaxies with pure exponential profiles (a total of
173) are widely distributed in morphological type and color, galaxies
with flattened exponential profiles (269) have, in the majority of the
cases, morphological types T in the range 2$<$T$<$8, i.e$.$ they are
truly spiral galaxies. In order to explain this behavior is necessary
to understand first what is the mechanism(s) behind the flattening of
the UV profiles.

In the spectro-photometric models of the evolution of disk galaxies of
Boissier \& Prantzos (2000; see also Boissier 2000), a similar
flattening in blue bands is obtained. The main reason for it is that
the rate the stars formed (i.e$.$ SFR) in the inner disk has been
higher than the infall of gas, leading to a progressive consumption of
the gas in these regions. In the outer parts, however, star formation
is less efficient and infall proceeds on longer timescales.  As a
result, the gas reservoir of the outer disk is not exhausted, and the
shape of the exponential profile is preserved (in adition, an
extinction gradient could enhance the difference between the inner
regions, metal and dust rich, and outer regions suffering low
metallicity and low extinction).

The dependence of the degree of flattening with the morphological type
found, with most galaxies showing flattened-exponential profiles
having types Sab-Sdm, is probably a consequence of the fact that (1)
early-type galaxies have already consumed the majority of their gas at
all radii, due to a high global star formation efficiency and low
current infall, and (2) late-type spirals, because of their current
large supply of gas and infall, still have enough gas to prevent its
consumption at all radii. Note also that in some very early type
spirals (S0/a and Sa types) the presence or a relatively bright bulge
might also difficult the detection of any flattening in the inner-disk
profile.

The models referred above use as parameters the circular velocity
(i.e$.$ total mass), and the spin parameter (i.e$.$ angular
momentum). For a fixed spin parameter, the degree of flatenning should
depend on mass since e.g$.$ the infall time-scale depends on the
mass. Indeed, at very low mass a modest flattening occurs, a more
visible one at intermediate mass, and no flatenning again in very
massive galaxies (where the gas has been consumed over the whole
galaxy) (Boissier 2000).  However, using the K-band absolute magnitude
as a tracer of the total mass of the system we found no difference
between the distribution of galaxies with or without flattening in
their profiles.  This disagreement with the naive expectation from the
models could be linked to the existence of the second parameter (at
fixed velocity, the flattening of the star formation rate is more
noticeable for smaller spin parameters), or more fundamental
differences between EF/Ef and EE galaxies, not yet included in models.
A more direct measure of the total mass and spin parameter, or
detailed modeling of these galaxies (or a sub-sample of them) could
help us to understand what makes the EF/Ef galaxies different from the
EE ones.

\section{Conclusions}
\label{conclusions}

We have presented an imaging Atlas of 1034 galaxies observed in two UV
bands by the GALEX satellite. From these we have derived surface
brightness and color profiles in the FUV \& NUV GALEX
bands. Asymptotic magnitudes and colors along with concentration
indices have also been obtained. A morphological classification of the
profiles is also carried out. Despite a small but non-negligible
excess of high-luminosity and paucity of low-luminosity spiral
galaxies (compared with the luminosity distribution of ellipticals
both in our and the NFGS samples) it is shown that this sample
adequately matches the distribution and full range of properties of
galaxies in the local Universe. We have augmented this data set with
corollary data from the optical (RC3), NIR (2MASS), and far-infrared
(IRAS). We emphasize here the special caution should be observed when
comparing these results with those derived from a volume-limited
sample. From a broad-based initial analysis of the UV properties of
this sample we conclude:

\begin{itemize}

\item The value of the integrated (FUV$-K$) color of galaxies provides 
an excellent criterion with which to discriminate
elliptical/lenticular galaxies from spirals and irregulars. The best
discrimination between these two classes of galaxies (quiescent vs$.$
star-forming) is achieved if a cut-off color (FUV$-K$)=8.8\,mag is
adopted. A reasonably good separation is also obtained by using a
(FUV$-$NUV) cut-off color at 0.9\,mag. These colors also allow for a
continuous distinction (although with a significant dispersion) of
spiral galaxies of different types.

\item Elliptical/lenticular galaxies with brighter FUV and $K$-band luminosities 
show bluer (FUV$-$NUV) colors than ellipticals with fainter
luminosities but redder (NUV$-K$) colors. This is true for ellipticals
galaxies specifically within the range of absolute magnitudes covered
by this Atlas (i.e$.$ M$_K$$<$$-$21\,mag). This behavior is probably a
consequence of luminous elliptical galaxies having stronger UV upturns
than their intermediate-mass counterparts (see Boselli et al$.$ 2005).

\item We do not find a large dispersion in the intrinsic (corrected for 
internal extinction) (FUV$-$NUV) colors of the spiral/irregular
galaxies in the Atlas ($\sigma_{\mathrm{(FUV-NUV)}_0}$=0.05\,mag)
neither a strong dependence of it with the galaxy
luminosity. Consequently, the variations in the observed (FUV$-$NUV)
colors with the luminosity or morphological type of the spiral and
irregular galaxies in the sample are plausibly due to variations in
the dust content (due for example to changes in metallicity) with
these magnitudes. In the case of the (FUV$-K$) color the star
formation history necessarily contributes to its dependence on
luminosity and morphological type.

\item The change in the observed (FUV$-$NUV) color with the TIR-to-FUV 
ratio also suggests that the attenuation law in these galaxies differs
from a pure Milky-Way extinction law. In particular, attenuation laws
with relatively steep FUV rise and no 2175\,\AA\ bump, like those
based on a SMC Bar extinction law or the Calzetti law in the case of
the most luminous objects, are favored.

\item A significant fraction (28\%) of the UV profiles show some degree 
of flattening in the inner regions. The galaxies showing this kind of
profiles belong to a relatively small range of optical morphological
types (compared with the pure-exponential profiles), 2$<$T$<$8, i.e$.$
they are all truly spiral galaxies. We interpret this as a consequence
of the high past SFR but comparatively low current gas infall rate in
the inner disks of spiral galaxies, leading to an efficient
consumption of the gas in these regions and, consequently, to a
flattening of the UV profiles compared with the outer disks, where the
gas supply is still abundant. This is, indeed, expected to be
particularly important in intermediate-type spirals.

\end{itemize}

The GALEX and corollary photometry data along with the profiles and UV
images of galaxies in the sample can be accessed through a dedicated
web page at\newline {\tt
http://nedwww.ipac.caltech.edu/level5/GALEX\_Atlas/}.

\acknowledgments

GALEX (Galaxy Evolution Explorer) is a NASA Small Explorer, launched in April 2003.
We gratefully acknowledge NASA's support for construction, operation,
and science analysis for the GALEX mission,
developed in cooperation with the Centre National d'Etudes Spatiales
of France and the Korean Ministry of 
Science and Technology. AGdP is partially financed by
the MAGPOP EU Marie Curie Research Training Network and the 
Spanish Programa Nacional de Astronom\'{\i}a y Astrof\'{\i}sica under grant AYA2003-01676.
We thank Cren Frayer and Olga Pevunova for preparing the online version of
the Atlas. We are also thankful to the referee for his/her valuable comments 
which helped to improve the paper. 

{\it Facilities:} \facility{GALEX}

\clearpage
\pagestyle{empty}
\input{agpaz_table1}
\clearpage
\input{agpaz_table2}

\clearpage
\input{agpaz_table3}
\clearpage
\input{agpaz_table4}
\clearpage
\pagestyle{plaintop}
\begin{figure}
\figurenum{1}
\epsscale{.9}
\plotone{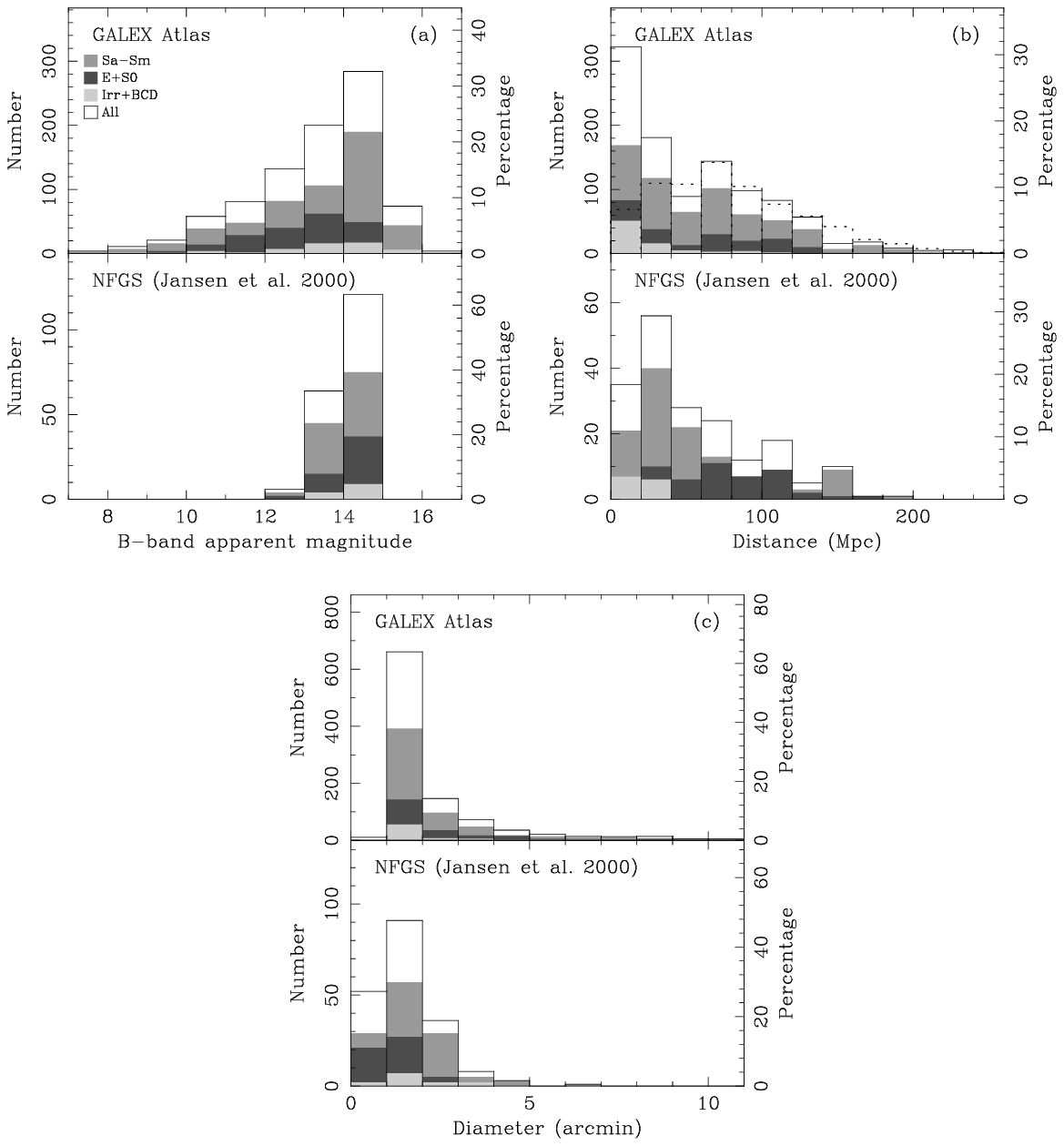}
\epsscale{1}
\caption{Comparison of the GALEX Atlas sample and the Nearby Field Galaxy Survey (NFGS) of Jansen et al$.$ (2000). Galaxies classified as ellipticals/lenticulars are shown in dark grey, spiral galaxies in grey, and irregulars and Blue Compact Dwarf (BCD) galaxies in light grey. The outlined solid-line histogram represents the distribution for all galaxies. (a) Distribution in $B$-band apparent magnitude. Note that the sharp cutoff in magnitude for the NFGS is due to the fact that this survey was extracted from the magnitude-limited CfA survey sample (Huchra et al$.$ 1983). (b) Distribution of distances in Mpc (see Section~\ref{sample.parameters} for a detailed description of how the distances to galaxies in the GALEX Atlas were determined). The broken-line histogram represents the distribution of 10,663 galaxies with measured redshifts in RC3. (c) Distribution of major-axis diameters in arcmin. Note that the size-limit for the serendipitous part of the GALEX Atlas was set to 1\,arcmin.\label{figure1}}
\end{figure}

\begin{figure}
\figurenum{2}
\plotone{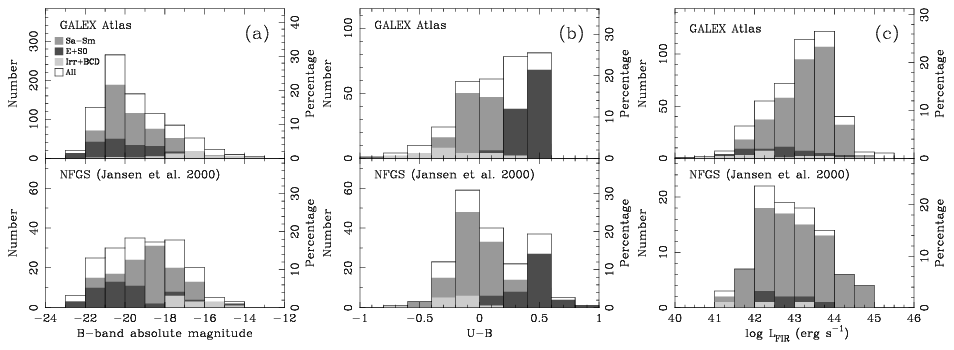}
\caption{Comparison of the properties of the GALEX Atlas sample and the NFGS. Color coding is as in Figure~1. (a) $B$-band absolute magnitude. Note that the GALEX Atlas sample covers the full range of properties of the galaxies in the Local Universe as described by the magnitude-limited NFGS sample. The distribution is quite similar in the case of elliptical/lenticular and irregular galaxies with only a moderate excess of intrinsically bright spirals in the case of the GALEX Atlas sample compared with field galaxies. It is worth noting that there are still almost three times more low-luminosity spiral galaxies in the GALEX Atlas than in the NFGS. The relative fractions of elliptical/spiral/irregular galaxies is also very similar between the GALEX Atlas sample and field galaxies (see text for details). (b) $U-B$ color. (c) FIR luminosity obtained from the IRAS 60 and 100\,$\mu$m fluxes using the recipe of Londsdale et al$.$ (1985).\label{figure2}}
\end{figure}

\figsetstart
\figsetnum{3}
\figsettitle{The GALEX Ultraviolet Atlas of Nearby Galaxies}

\figsetend

\begin{figure}
\figurenum{3}
\epsscale{0.93}
\plotone{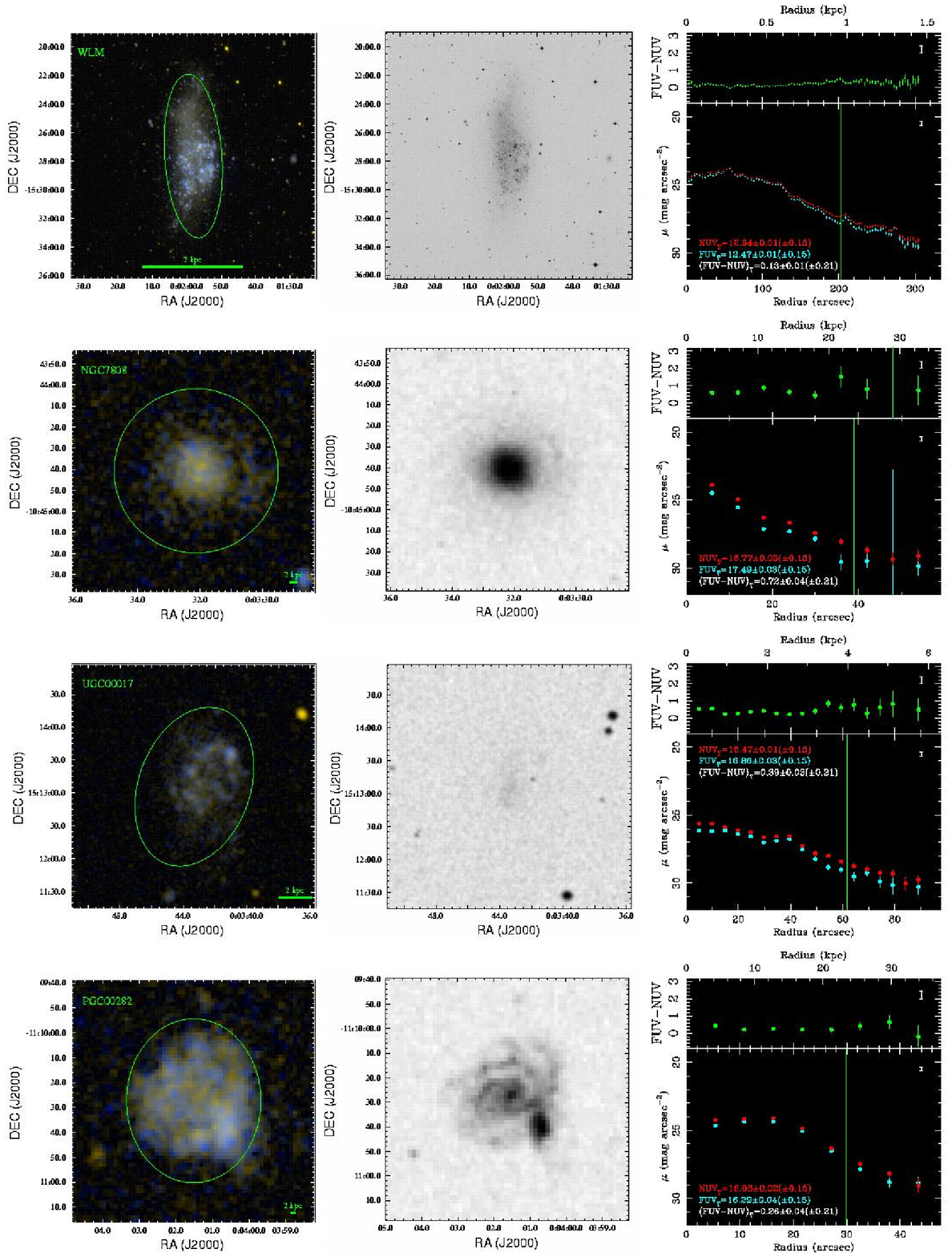}
\caption{False-color GALEX images (left), DSS-1 images (center), surface brightness and color profiles (right) of the galaxies in the Atlas (see text for details). All panels are available online at {\tt http://nedwww.ipac.caltech.edu/level5/GALEX\_Atlas/}.\label{figure3}}
\epsscale{1.0}
\end{figure}

\clearpage
\epsscale{0.8}
{\plotone{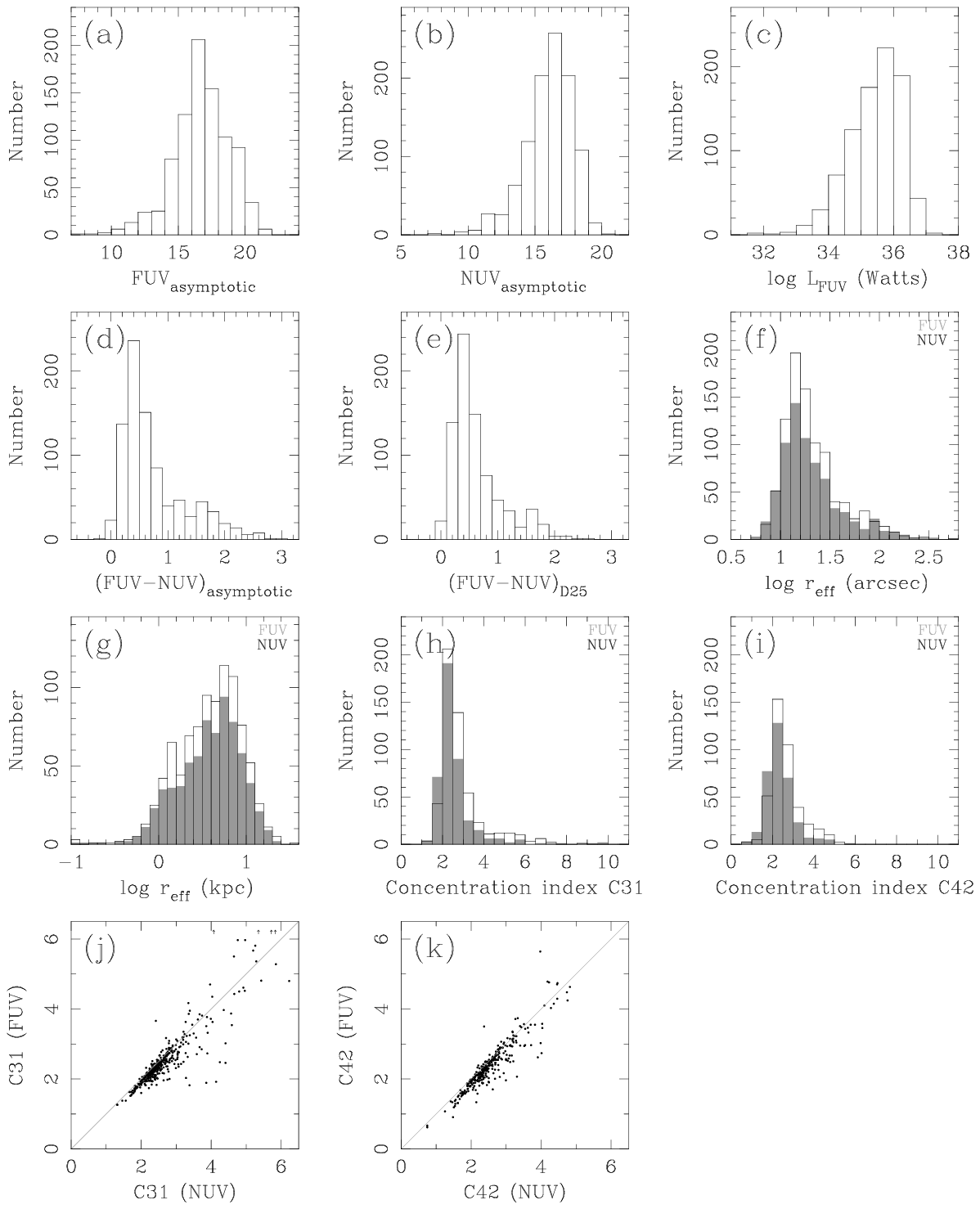}}
\epsscale{1.0}
\clearpage
\begin{figure}
\figurenum{4}
\caption{UV properties of the galaxies in the Atlas. {\bf a)} Frequency histogram of apparent asymptotic FUV magnitudes (AB scale). {\bf b)} Apparent asymptotic NUV magnitudes. {\bf c)} FUV luminosity in Watts computed as $\nu$\,F$_\nu$ (see Buat et al$.$ 2005). {\bf d)} (FUV$-$NUV) color. {\bf e)} (FUV$-$NUV) color inside the D25 ellipse. {\bf f)} Effective radius (in arcsec) of the galaxies in the FUV (gray-shaded histogram) and NUV (outlined histogram). {\bf g)} The same with the radius in kpc. {\bf h)} FUV (gray-shaded histogram) and NUV (outlined histogram) C31 concentration index. {\bf i)} The same for the C42 concentration index. {\bf j)} Comparison of the FUV and NUV C31 concentration indices. {\bf k)} The same for the C42 concentration index.\label{figure4}}
\epsscale{1.0}
\end{figure}

\clearpage
\epsscale{0.8}
{\plotone{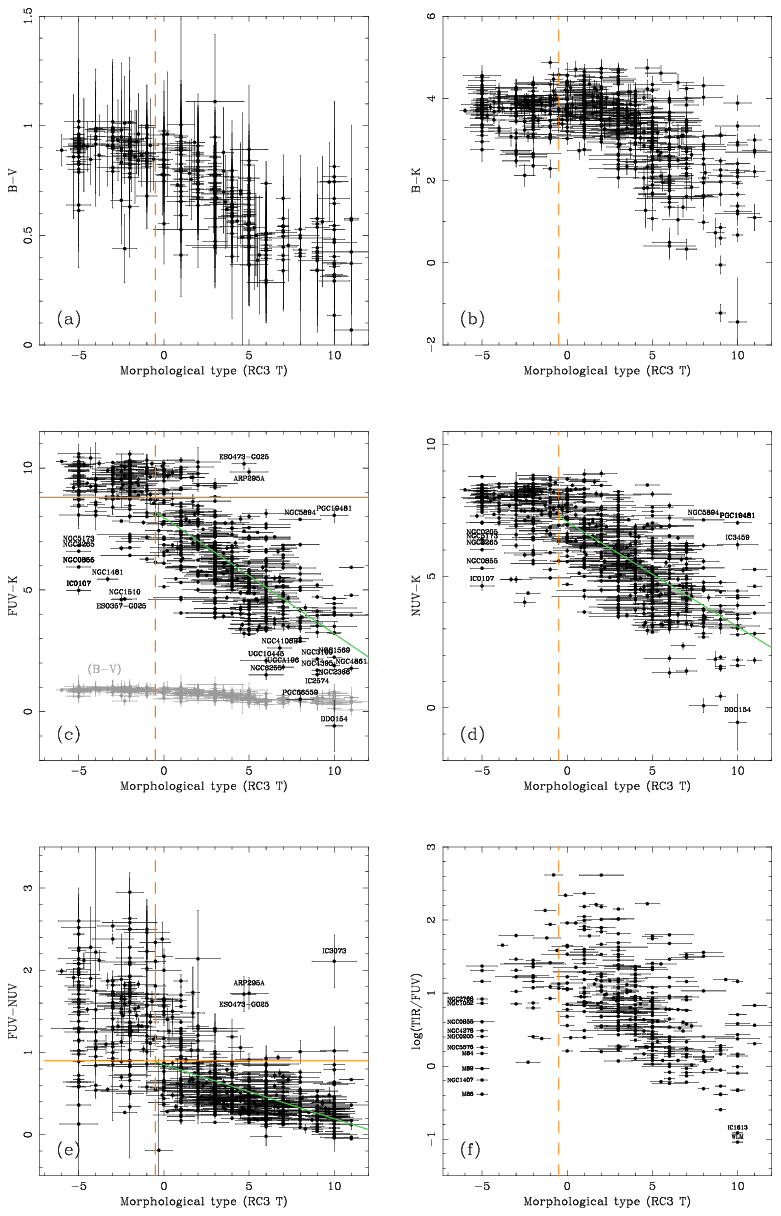}}
\clearpage
\begin{figure}
\figurenum{5}
\caption{Variation in the observed colors and total-infrared (TIR) to FUV ratio of the galaxies in the Atlas with the morphological type (T). {\bf a)} ($B-V$) versus the morphological type for elliptical/lenticular (T$<-$0.5), spiral ($-$0.5$\leq$T$<$9.5), and irregular/compact galaxies (T$\geq$9.5). The separation between elliptical/lenticular and the rest is shown by a vertical dashed line. {\bf b)} The same for ($B-K$). Note the small segregation in color between the different types when the ($B-V$) or ($B-K$) colors are used. {\bf c)} The same for (FUV$-K$). The segregation between ellipticals/lenticulars and spirals (horizontal solid line) and even between different kind of spiral galaxies is now remarkable. For comparison purposes we show (in the same scale) the range in ($B-V$) color span by the galaxies in the sample (see panel {\bf a}). {\bf d)} The same for (NUV$-K$). {\bf e)} The same for the (FUV$-$NUV) color. Note that FUV and NUV magnitudes are in AB scale and optical and NIR magnitudes are in the Vega system. Green lines represent the best linear fit to the data for types T=$-$0.5 or later (i.e$.$ spiral galaxies). {\bf f)} The same for the TIR-to-FUV ratio (see Buat et al$.$ 2005).\label{figure5}}
\epsscale{1.0}
\end{figure}

\clearpage
\epsscale{0.8}
{\plotone{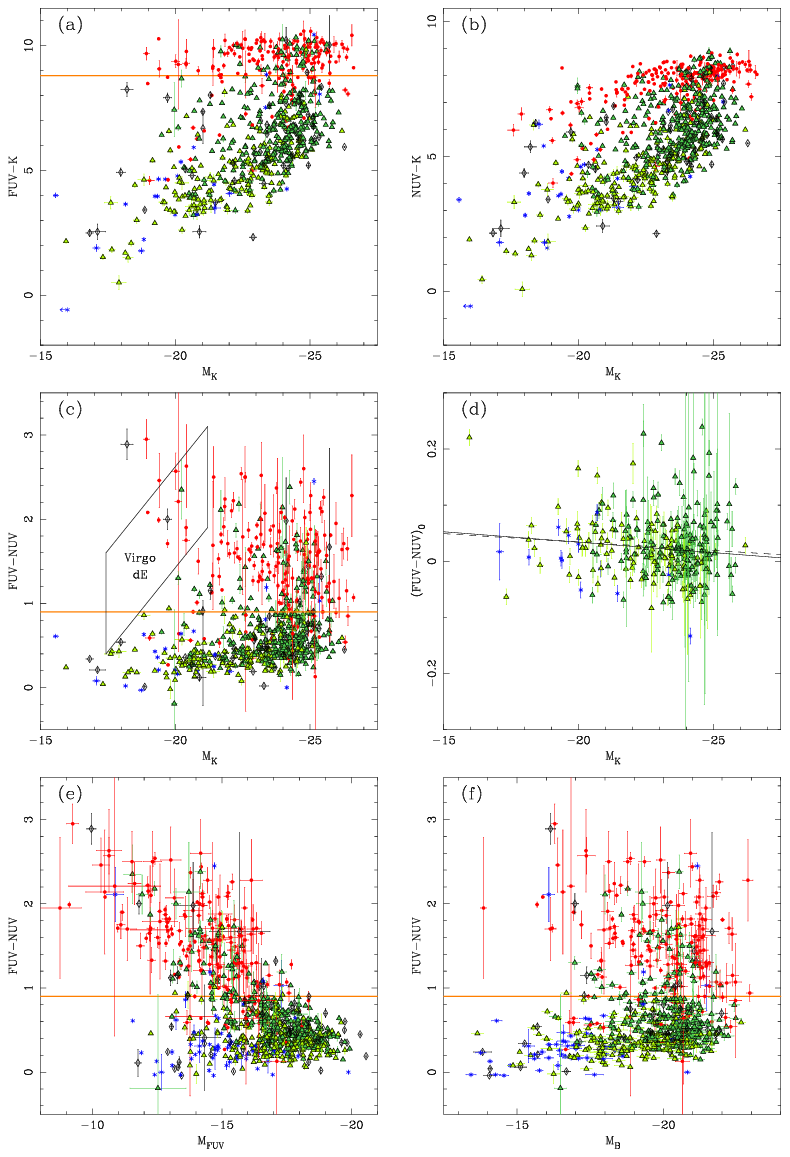}}
\epsscale{1.0}
\clearpage
\begin{figure}
\figurenum{6}
\caption{Color-magnitude diagrams (CMD) of the Atlas galaxies. 
Red dots are elliptical/lenticular galaxies, dark green triangles are early-type spirals (T$<$5), light green triangles are late-type spirals (T$\geq$5), blue
asterisks are irregular and compact galaxies, and black diamonds are
galaxies currently lacking morphological classification. {\bf a)}
(FUV$-K$)-M$_K$ CMD. Spiral and irregular galaxies show a systematic
bluing as we move to galaxies of lower mass. Elliptical/lenticular galaxies, on
the other hand, show a very small change in the (FUV$-K$) color with
the $K$-band absolute magnitude (i.e$.$ stellar mass) of the
galaxy. {\bf b)} (NUV$-K$)-M$_K$ CMD. In this case, however, lower
mass ellipticals are also systematic bluer than more massive
ones. {\bf c)} (FUV$-$NUV)-M$_K$ CMD. This plot shows that the
behavior observed in the elliptical galaxies in previous diagrams
seems to be consequence of a much fainter UV upturn (best traced by
the FUV-NUV color) in low-luminosity ellipticals than in massive
ones. In this plot we show the position occupied by dwarf elliptical
galaxies in Virgo (Boselli et al$.$ 2005). Dwarf elliptical galaxies
fainter than M$_K$$<$$-$21\,mag start to show the effects of recent
star formation both on their (FUV$-$NUV) and UV-optical colors (see
Boselli et al$.$ 2005 for more details). {\bf d)}
(FUV$-$NUV)$_0$-M$_K$ CMD. The (FUV$-$NUV)$_0$ color is corrected for
internal extinction using the relation between the total-infrared
(TIR) to FUV ratio and the extinction in the FUV and NUV bands given
by Buat et al$.$ (2005). Only spiral and irregular/compact galaxies
are used in this plot. Solid (dashed) line represents the best
weighted (non-weighted) fit to the data. The narrow distribution in
extinction-corrected UV slopes indicates that the tendency seen in the
(FUV$-$NUV)-M$_K$ CMD shown above for spiral galaxies is a direct
consequence of the change in the amount of dust with the luminosity of
the galaxy. {\bf e)} (FUV$-$NUV)-M$_{\mathrm{FUV}}$ CMD. {\bf f)}
(FUV$-$NUV)-M$_B$ CMD. These two latter diagrams show a similar
behavior to that shown in panel {\bf c}.The high-luminosity end of the
sample in the FUV is clearly dominated by spiral galaxies with a very
narrow distribution in observed (FUV$-$NUV) color.
\label{figure6}}
\end{figure}

\begin{figure}
\figurenum{7}
\epsscale{1.0}
\plottwo{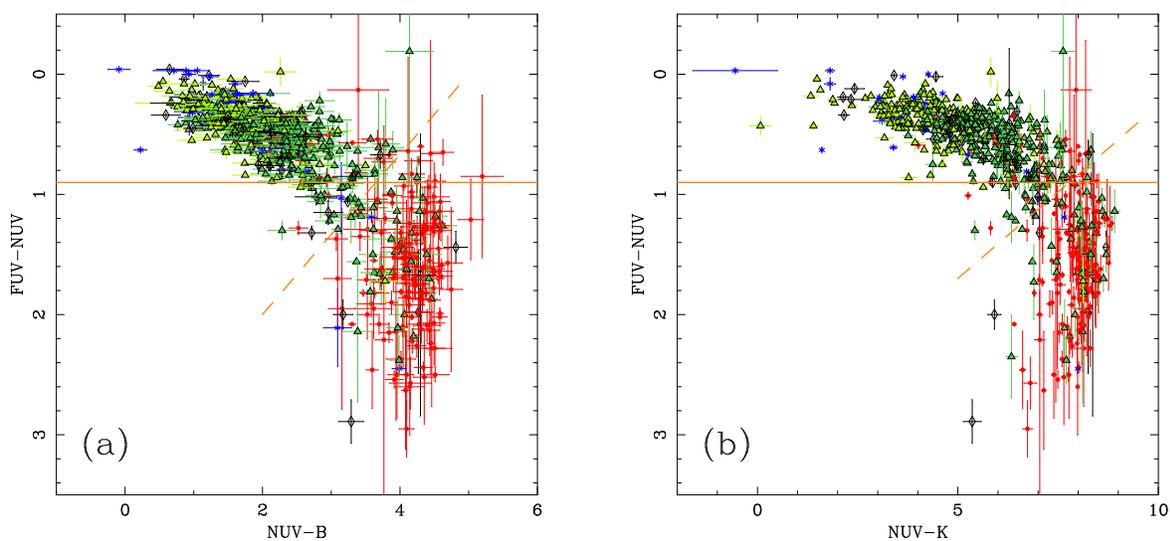}{agpaz_f7b.ps}
\caption{Color-color diagrams of the galaxies in the Atlas. {\bf a)} (FUV$-$NUV)-(NUV$-B$) color-color diagram. {\bf b)} (FUV$-$NUV)-(NUV$-K$) color-color diagram. Symbols have the same meaning as in Figure~6. Lines in this plot represent various criteria proposed to separate elliptical/lenticular galaxies from spirals (see text for details). Note that, in order to keep with the stellar convention, the (FUV$-$NUV) axis has been flipped and red (FUV$-$NUV) colors are now plotted at the bottom of the figure.\label{figure7}}
\epsscale{1.0}
\end{figure}

\clearpage
\begin{figure}
\figurenum{8}
\plottwo{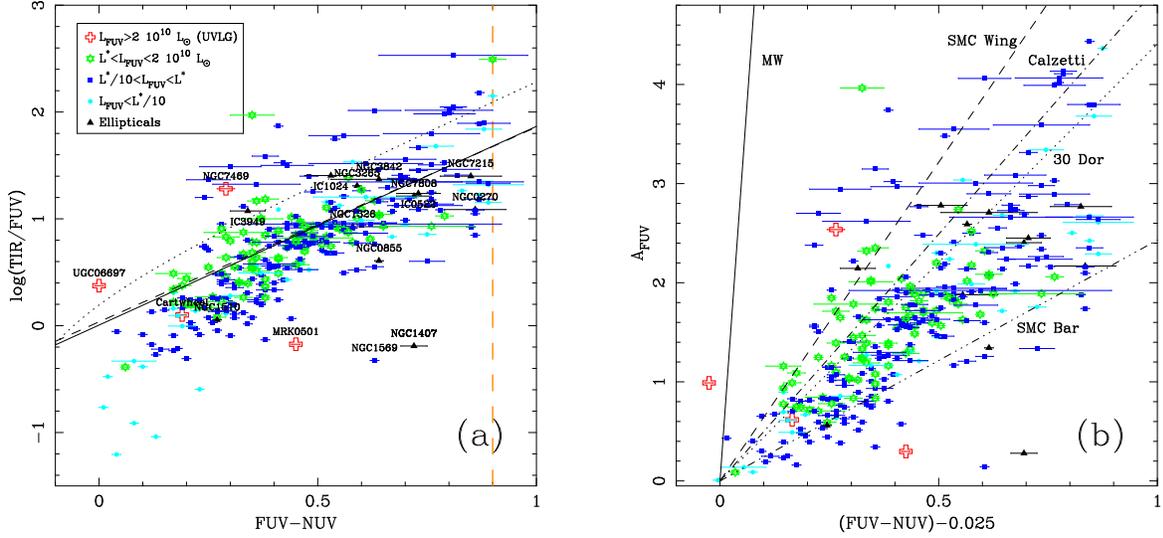}{agpaz_f8b.ps}
\caption{{\bf a)} IRX-beta relation. The vertical long dashed-line represents the cutoff in (FUV$-$NUV) color used to select the sub-sample of galaxies used to study the relation between the TIR-to-FUV ratio and the slope of the UV. This selection criterion guarantees that in the galaxies considered both the infrared and the UV emission are in the most part associated with the presence of recent star formation activity. The dotted line represents the relation derived by using a sample of starburst galaxies (Kong et al$.$ 2004; Meurer et al$.$ 1999). The best fit to the whole set of data is shown by a solid line. The best fit obtained excluding the lowest luminosity galaxies (dots) is shown by a dashed line. Symbols are coded by UV luminosity. Galaxies with higher UV luminosities (green stars) seem to be located somewhat closer to the relation derived for starburst galaxies that fainter objects (blue squares). Triangles correspond to the elliptical galaxies in the sample. Note that most of the ellipticals with (FUV$-$NUV)$<$0.9 are known to have some degree of residual star formation. {\bf b)} A$_{\mathrm{FUV}}$ versus (FUV$-$NUV)$-$0.025. The latter term is equivalent to A$_{\mathrm{FUV}}$$-$A$_{\mathrm{NUV}}$ if an intrinsic (FUV$-$NUV)=0.025\,mag is assumed for all star-forming galaxies in the sample (see Section~\ref{results.cmd}). The lines drawn correspond to the total-to-selective extinction in the UV expected for different extinction laws (MW, solid line; LMC 30~Doradus, dotted line; SMC Wing, dashed line; SMC Bar, dot-dot-dot-dashed line) and the attenuation law of Calzetti et al$.$ (1994, dot-dashed line). The $R_V$ values adopted for each of these laws are given in the text. Note that the inclusion of scattering would result in steeper relations between A$_{\mathrm{FUV}}$ and A$_{\mathrm{FUV}}$$-$A$_{\mathrm{NUV}}$ than those shown here. Therefore, an attenuation law based on the SMC Bar extinction law seems to be favored by these results. High-UV-luminosity galaxies are still adequately represented by the Calzetti attenuation law.\label{figure8}}
\epsscale{1.0}
\end{figure}

\begin{figure}
\figurenum{9}
\epsscale{1.0}
\plotone{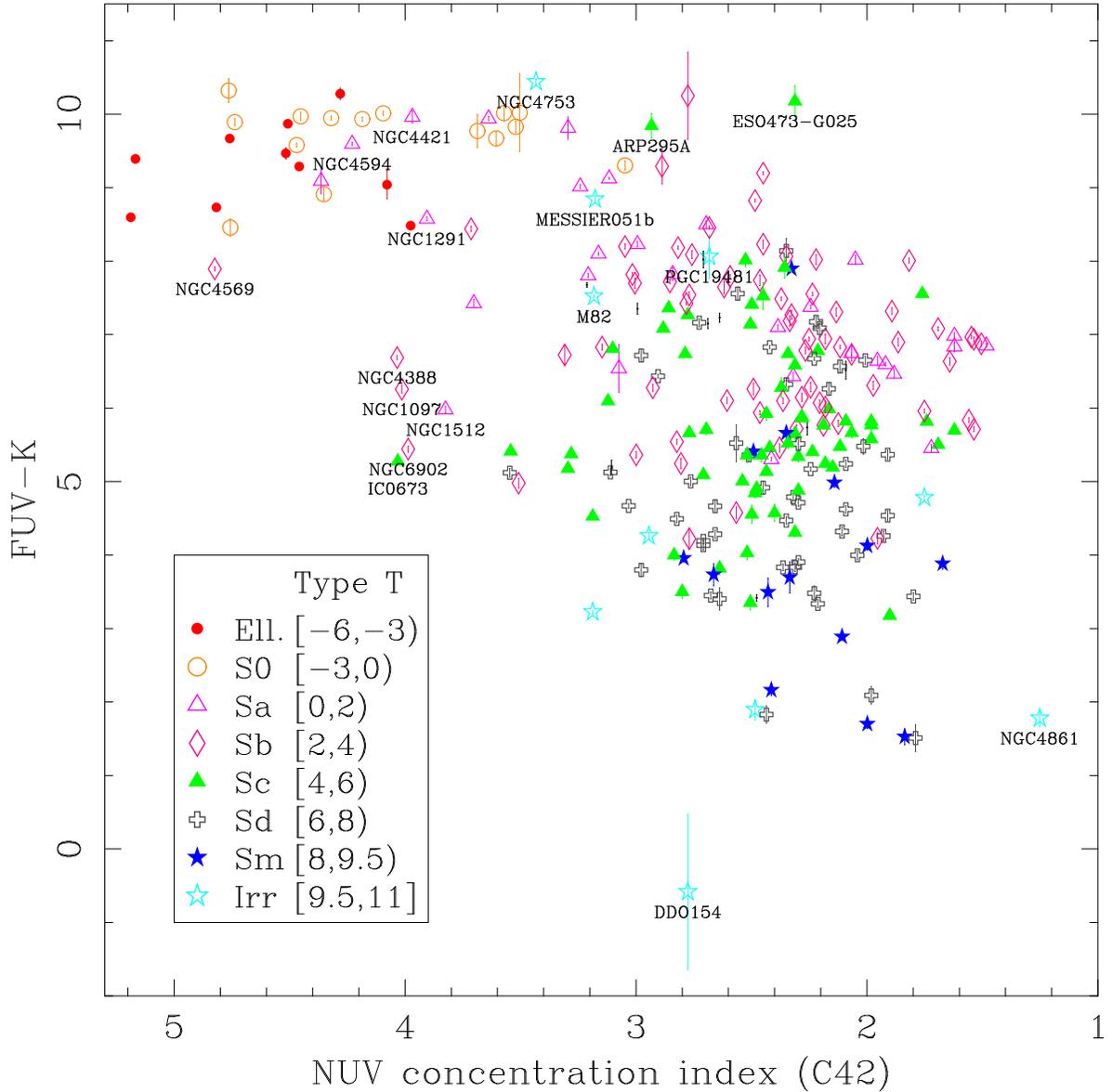}
\caption{(FUV$-K$) color versus the concentration index C42 in the NUV. Symbols are coded by morphological type. Although the galaxies are better segregated in (FUV$-K$) color than in concentration index, the value of C42 can be used improve to the discrimination between ellipticals and lenticulars and between these and some early-type spirals.\label{figure9}}
\epsscale{1.0}
\end{figure}

\begin{figure}
\figurenum{10}
\epsscale{1.0}
\plotone{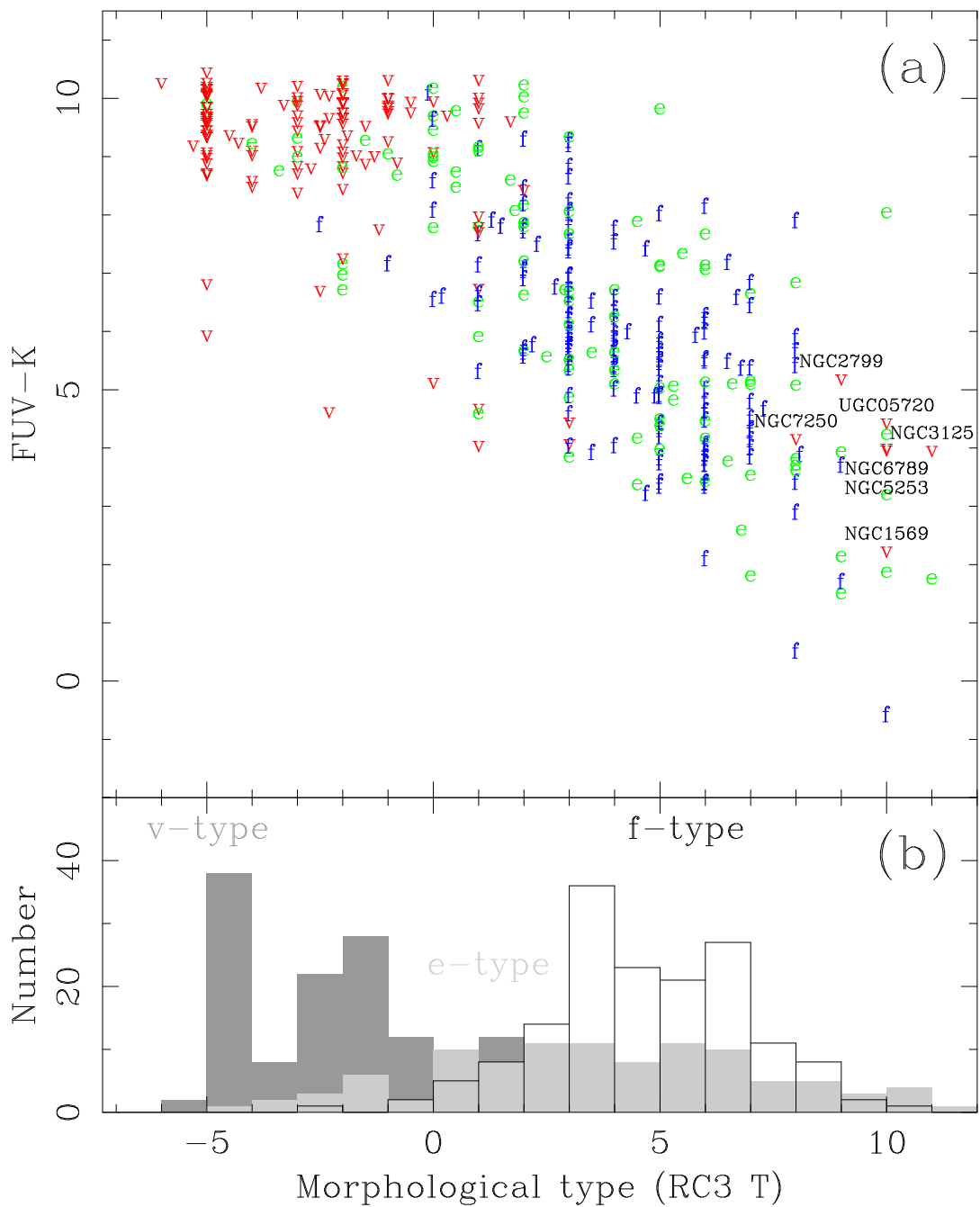}
\caption{{\bf a)} (FUV$-K$) color versus the morphological type. The symbols are coded by letters that represent the morphology of their UV profiles: {\bf v} for galaxies following a de Vaucouleurs R$^{1/4}$ profile, {\bf e} for galaxies with pure exponential profiles, and {\bf f} for galaxies with exponential profiles in the outer regions and a flattened profile inside. {\bf b)} Morphological-type distribution for each class of UV profile.\label{figure10}}
\epsscale{1.0}
\end{figure}

\end{document}

%% file: agpaz_table1.tex
\clearpage
\begin{deluxetable}{lccrrrrcrclcl}
\rotate
\tabletypesize{\scriptsize}
\tablecaption{GALEX Atlas sample\label{table1}}
\tablewidth{0pt}
\tablehead{
\colhead{Object Name} & \colhead{RA$_{2000}$}   & \colhead{DEC$_{2000}$}   & \colhead{2$\times$A} & \colhead{2$\times$B}  & \colhead{PA} & \multicolumn{3}{c}{distance} & \colhead{E($B-V$)} & \colhead{Morphological} & \colhead{T} & \colhead{Spectral}\\
\colhead{}            &  \colhead{(h:m:s)}    & \colhead{(d:m:s)}      &  \colhead{(arcmin)}   & \colhead{(arcmin)}    & \colhead{(deg)} & \colhead{(Mpc)} & \colhead{DM} & \colhead{ref.} & \colhead{(mag)} & \colhead{Type} & \colhead{Type} & \colhead{Type}\\
\colhead{(1)} & \colhead{(2)} & \colhead{(3)} & \colhead{(4)} & \colhead{(5)} &\colhead{(6)} &\colhead{(7)} &\colhead{(8)} &\colhead{(9)} & \colhead{(10)} &\colhead{(11)} &\colhead{(12)} & \colhead{(13)}}
\startdata
WLM             &  00 01 58.2  & $-$15 27 39.3  &  11.5  &   4.0  &   4      & 0.98  & 24.96 &  2  & 0.04  & IB(s)m          &  10.0$\pm$0.3   &             \\      
NGC~7808         &  00 03 32.1  & $-$10 44 40.8  &   1.3  &   1.3  & \nodata  & 124  & 35.48 &  1  & 0.04  & (R')SA0*0*:     &  -2.0$\pm$1.2   &        Sy1? \\       
UGC~00017        &  00 03 43.3  & $+$15 13 06.0  &   2.5  &   1.7  & 160      &  13  & 30.62 &  1  & 0.05  & Sm:             &   9.3$\pm$0.6   &             \\       
PGC~00282        &  00 04 01.5  & $-$11 10 27.3  &   1.1  &   0.9  &   0      & 161  & 36.04 &  1  & 0.04  & SB(rs)c:        &   5.0$\pm$1.0   &             \\       
NGC~0024         &  00 09 56.5  & $-$24 57 47.3  &   5.8  &   1.3  &  46      & 8.2  & 29.57 &  3  & 0.02  & SA(s)c          &   5.0$\pm$0.3   &             \\       
UGC~00128        &  00 13 50.9  & $+$35 59 39.0  &   2.1  &   1.7  &  65      &  67  & 34.13 &  1  & 0.06  & Sdm             &   8.0$\pm$0.7   &             \\       
NGC~0055         &  00 14 53.6  & $-$39 11 47.9  &  32.4  &   5.6  & 108      & 2.0  & 26.51 &  4  & 0.01  & SB(s)m: sp      &   9.0$\pm$0.7   &             \\       
ARP~256~NED02     &  00 18 50.1  & $-$10 21 41.8  &   1.1  &   0.8  &   0      & 116  & 35.32 &  1  & 0.04  & SB(s)c pec      &   4.5$\pm$0.6   &         HII \\
\enddata
\end{deluxetable}

%% file: agpaz_table2.tex
\clearpage
\begin{deluxetable}{lcrlccccccr}
\rotate
\tabletypesize{\scriptsize}
\tablecaption{GALEX observations\label{table2}}
\tablewidth{0pt}
\tablehead{
\colhead{Object Name} & \colhead{Date}   & \colhead{Exposure}   & \colhead{Tile} & \multicolumn{3}{c}{FUV sky background}  & \multicolumn{3}{c}{NUV sky background} & \colhead{notes}\\
\colhead{}            &  \colhead{Observed}    & \colhead{(sec)}    &  \colhead{}   & \colhead{mean}    & \colhead{$<$$\sigma$$>$} & \colhead{$\sigma$(mean)} & \colhead{mean}    & \colhead{$<$$\sigma$$>$} & \colhead{$\sigma$(mean)} & \\
\colhead{}            &  \colhead{}    & \colhead{}    &  \colhead{}   & \colhead{(counts/s)}    & \colhead{(counts/s)} & \colhead{(counts/s)} & \colhead{(counts/s)}    & \colhead{(counts/s)} & \colhead{(counts/s)} & \\
\colhead{(1)} & \colhead{(2)} & \colhead{(3)} & \colhead{(4)} & \colhead{(5)} &\colhead{(6)} &\colhead{(7)} &\colhead{(8)} &\colhead{(9)} & \colhead{(10)} &\colhead{(11)}}
\startdata
WLM             & 2003-10-23 &  1442 &  NGA-WLM                       & 2.657E-04 & 4.595E-04 & 1.133E-05 & 2.968E-03 & 1.681E-03 & 8.973E-06 &                 \\
NGC~7808         & 2003-09-17 &   592 &  MISDR1-29571-0650             & 2.403E-04 & 6.984E-04 & 3.168E-05 & 3.119E-03 & 2.688E-03 & 1.062E-05 &                 \\
UGC~00017        & 2004-08-26 &  1705 &  MISDR2-28657-0750             & 4.114E-04 & 4.988E-04 & 1.217E-05 & 3.018E-03 & 1.500E-03 & 9.886E-06 &      (1)        \\
PGC~00282        & 2003-09-17 &   592 &  MISDR1-29571-0650             & 2.228E-04 & 7.111E-04 & 2.995E-06 & 3.116E-03 & 2.730E-03 & 1.126E-04 &                 \\
NGC~0024         & 2003-09-16 &  1602 &  NGA-NGC0024                   & 2.434E-04 & 4.232E-04 & 3.274E-05 & 2.789E-03 & 1.537E-03 & 8.181E-05 &                 \\
UGC~00128        & 2003-09-09 &  1405 &  NGA-UGC0128                   & 5.992E-04 & 8.596E-04 & 1.123E-05 & 3.526E-03 & 1.794E-03 & 4.927E-05 &                 \\
NGC~0055         & 2003-09-14 &  1511 &  NGA-NGC0055                   & 2.564E-04 & 4.457E-04 & 5.672E-05 & 2.503E-03 & 1.512E-03 & 1.213E-04 &                 \\
ARP~256~NED02     & 2003-09-19 &  1429 &  MISDR1-29523-0652             & 2.269E-04 & 4.122E-04 & 7.915E-06 & 3.016E-03 & 1.627E-03 & 2.816E-06 &                 \\
\enddata
\tablecomments{{\rm Log} of GALEX observations. Col$.$ (1): Galaxy name. Col$.$ (2): Date the galaxy was observed. Col$.$ (3): Total exposure time in seconds for each of the GALEX bands. Col$.$ (4): Tile name assigned by the GALEX mission. It includes the name of the survey as part of which the galaxy was observed and the name of the targetted field. Col$.$ (5): Mean sky background of the FUV image (in counts per second). Col$.$ (6): Mean standard deviation of the sky in the FUV image measured by averaging the standard deviation within several regions around the position of the object. Col$.$ (7): Standard deviation of the mean values of the sky measured in these regions. Cols$.$ (8,9,10): The same for the NUV image. Col$.$ (11): Notes for individual objects: (1) Nominal galaxy position angle changed, (2) nominal galaxy coordinates changed, (3) special pipeline processing, (4) NUV-only, (5) very shallow FUV image, (6) very bright star near or on top of the galaxy, (7) nominal galaxy size changed, (8) crowded field, uncertain photometry.
}
\end{deluxetable}